\pdfoutput=1
\documentclass[journal]{IEEEtran}
\usepackage[cmex10]{amsmath}

\usepackage{subfigure}

\usepackage{setspace}
\usepackage{blindtext}
\usepackage{stackrel}
\usepackage{etoolbox}
\usepackage{enumerate}
\usepackage{amssymb}
\usepackage{tikz, pgfplots}
\usepgfplotslibrary{groupplots}
\usepgfplotslibrary{external}
\usepackage{color,soul}
\usetikzlibrary{calc}
\usetikzlibrary{arrows}
\usetikzlibrary{arrows,positioning} 
\usetikzlibrary{decorations.markings}
\usetikzlibrary{shapes}
\tikzset{
    >=stealth',
    punkt/.style={
           rectangle,
           rounded corners,
           draw=black, very thick,
           text width=6.5em,
           minimum height=2em,
           text centered},
    pil/.style={
           ->,
           thick,
           shorten <=2pt,
           shorten >=2pt,}
    pir/.style={
           <-,
           thick,
           shorten <=2pt,
           shorten >=2pt,}
}
\definecolor{KeynoteRed}{rgb}{.678,.051, .051}
\definecolor{KeynoteBlue}{rgb}{0.008, 0.443, 0.60}
\definecolor{KeynoteLightblue}{rgb}{.635, .914, .973}
\definecolor{KeynoteYellow}{rgb}{0.859, 0.584, 0.212}
\definecolor{KeynoteYellow}{rgb}{0.859, 0.584, 0.212}
\definecolor{KeynoteSlate}{rgb}{0.239, 0.271, 0.322}
\definecolor{KeynoteGray}{rgb}{0.498, 0.529, 0.529}
\definecolor{KeynoteGreen}{rgb}{0.18, 0.5, 0.08}
\definecolor{KeynoteTextGray}{rgb}{0.325, 0.325, 0.325}
\definecolor{KeynoteLightGray}{rgb}{0.706, 0.706, 0.706}
\definecolor{KeynoteBlueGray}{rgb}{0.471, 0.533, 0.620}
\definecolor{ECEpurple}{rgb}{.169, .18, .455}
\definecolor{ECEcyan}{rgb}{.41, .62, .72}
\definecolor{ECEgray}{rgb}{.788, .827, .859}
\definecolor{ECEblueGray}{rgb}{61.2, 70.6, 70.6}
\definecolor{ECEblueGray}{rgb}{61.2, 70.6, 70.6}
\definecolor{RiceBlue}{rgb}{0, .14, .41}

\newtoggle{ZChanModel}

\usepackage{cite}
\usepackage{ifthen,verbatim}
\newif\ifcomments

\newif\ifPlotTempData

\PlotTempDatafalse
\commentsfalse
\newcommand{\txDims}{effective antennas}

\newcommand{\dataDir}{DocData}

\newcommand{\bigArray}{many-antenna}

\newcommand{\softNull}{SoftNull}

\newcommand{\numRx}{M_{\mathsf{Rx}}}

\newcommand{\numTx}{M_{\mathsf{Tx}}}

\newcommand{\numUplink}{K_\mathsf{Up}}

\newcommand{\numDownlink}{K_\mathsf{Down}}

\newcommand{\numTotal}{M}

\newcommand{\upChannel}{\boldsymbol{H}_\mathsf{up}}

\newcommand{\downChannel}{\boldsymbol{H}_\mathsf{Down}}

\newcommand{\interuserChannel}{\boldsymbol{H}_\mathsf{Usr}}

\newcommand{\selfChannel}{\boldsymbol{H}_\mathsf{Self}}

\newcommand{\selfPrecoder}{\boldsymbol{P}_\mathsf{Self}}

\newcommand{\precoder}{\boldsymbol{P}}

\newcommand{\downPrecoder}{\boldsymbol{P}_\mathsf{Down}}

\newcommand{\herm}[1]{#1^\mathrm{H}}

\newcommand{\downSyms}{\boldsymbol{s}_\mathsf{Down}}

\newcommand{\bsRx}{\boldsymbol{y}_\mathsf{Up}}

\newcommand{\bsNoise}{\boldsymbol{z}_\mathsf{Up}}

\newcommand{\usrRx}{\boldsymbol{y}_\mathsf{Down}}

\newcommand{\usrNoise}{\boldsymbol{z}_\mathsf{Down}}

\newcommand{\usrTx}{\boldsymbol{x}_\mathsf{Up}}

\newcommand{\bsTx}{\boldsymbol{x}_\mathsf{Down}}

\newcommand{\identity}[2]{\boldsymbol{I}_{#1\times#2}}

\newcommand{\singLeft}{\boldsymbol{U}}

\newcommand{\singRight}{\boldsymbol{V}}

\newcommand{\singVals}{\boldsymbol{\Sigma}}

\newcommand{\singVecRight}[1]{\boldsymbol{v}^{(#1)}}

\newcommand{\eigval}{{\lambda}}

\newcommand{\eigvec}{\boldsymbol{v}}

\DeclareMathOperator{\trace}{Tr}

\DeclareMathOperator{\argmin}{argmin}

\newcommand{\userDim}{D_\mathsf{Tx}}

\newcommand{\mimoLayer}{standard MU-MIMO}

\newcommand{\softNullLayer}{self-interference reduction}

\newcommand{\effDownChannel}{\boldsymbol{H}_\mathsf{Eff}}

\newcommand{\lagMult}{\boldsymbol{\Lambda}}

\definecolor{DarkBlue}{rgb}{0, 0, 0.6}
\newcommand{\revision}[1]{#1}
\newcommand{\revisionTwo}[1]{#1}

\ifcomments
	\newcommand{\evanNote}[1]{{\color{blue} {\emph{#1}}}}
	\newcommand{\ashuNote}[1]{{\color{red} {\emph{#1}}}}
	\newcommand{\clay}[1]{{\color{KeynoteYellow} {\emph{#1}}}}
	\newcommand{\lin}[1]{{\color{KeynoteGreen} {\emph{#1}}}}
\else
	\newcommand{\evanNote}[1]{}
	\newcommand{\ashuNote}[1]{}
	\newcommand{\clay}[1]{}
	\newcommand{\lin}[1]{}
\fi
	\newcommand{\evanOldNote}[1]{}
	\newcommand{\ashuOldNote}[1]{}
	\newcommand{\clayOld}[1]{}
	\newcommand{\linOld}[1]{}

\newcommand{\dB}{\ \mathrm{dB}}
\newcommand{\dBm}{\ \mathrm{dBm}}

\newcommand{\SINR}{{\sf SINR}}

\newcommand{\scheme}{{\sf S}}
\newcommand{\HD}{{\sf HD}}
\newcommand{\IFD}{{\sf IdealFD}}
\newcommand{\SN}{{\sf SoftNull}}

\newcommand{\DownSINR}[1]{\SINR_{\rm Down}^{(#1)}}
\newcommand{\UpSINR}[1]{\SINR_{\rm Up}^{(#1)}}

\newcommand{\DownRate}[1]{{R_{\rm Down}^{(#1)} } }
\newcommand{\UpRate}[1]{{R_{\rm Up}^{(#1)} } }
\newcommand{\upFraction}[1]{\alpha_{\rm Up}^{(#1)} }
\newcommand{\downFraction}[1]{\alpha_{\rm Down}^{(#1)} }

\begin{document}

\title{\revision{\softNull:  Many-Antenna Full-Duplex Wireless via Digital Beamforming}}

\author{Evan~Everett, Clayton Shepard, Lin Zhong, and Ashutosh~Sabharwal
\thanks{All authors are with Rice University Department of Electrical and Computer Engineering (email: evan.everett, cws, lzhong, ashu@rice.edu).}
\thanks{This work was partially supported by NSF Grants CNS 0923479, CNS 1012921 and CNS 1161596 and NSF Graduate Research Fellowship 0940902.}}
%

\maketitle

\begin{abstract}
In this paper, we present and study a \revision{digital-controlled} method, called \softNull, to enable full-duplex in many-antenna systems. Unlike most designs that rely on analog cancelers to suppress self-interference, \softNull\ 
relies on digital transmit beamforming to reduce self-interference.
\softNull{} does not attempt to perfectly null self-interference, but instead seeks to reduce self-interference  sufficiently to prevent swamping the receiver's dynamic range. Residual  self-interference is then cancelled digitally by the receiver. We evaluate the performance of \softNull{} using measurements from a 72-element antenna array in both indoor and outdoor environments. We find that \softNull{} can significantly outperform half-duplex for small cells operating in the \bigArray{} regime, where the number of antennas is many more than the number of users served simultaneously.

\end{abstract}


%
\IEEEpeerreviewmaketitle

\section{Introduction}
\label{sec:introduction}


Full-duplex wireless communication, in which transmission and reception occur at the same time and in the same frequency band, has the potential to as much as double the spectral efficiency of traditional half-duplex systems. 
The main challenge to full-duplex is self-interference: a node's transmit signal generates high-powered interference to its own receiver. 
Research over the last ten years \cite{Bliss07SimultTX_RX, Choi10FullDuplex,Duarte10FullDuplex, Duarte11FullDuplex, Jain2011RealTimeFD, Duarte2012FullDuplexWiFi, sigcomm13fullDuplexRadios, Riihonen11FDMIMO, Sahai13PhaseNoise, Aryafar12MIDU} has shown that full-duplex operation may be feasible for small cells,  and the key enabler has been \emph{analog cancellation} of the self-interference in addition to digital cancellation. 
Analog cancellation has been considered a necessary component of a full-duplex system, to avoid self-interference from overwhelming the dynamic range of the receiver electronics, and swamping the much weaker intended signal (see Appendix~\ref{sec:dynamicRangeExample} for a detailed explanation of dynamic range and its impact full-duplex operation).

Many analog cancellation designs have been proposed for single-antenna \cite{Jain2011RealTimeFD, sigcomm13fullDuplexRadios} and dual-antenna \cite{Choi10FullDuplex, Duarte10FullDuplex, Duarte11FullDuplex} full-duplex systems. 
However, current wireless base stations use many antennas (up to 8 in LTE Release 12~\cite{IEEE-LTE-Rel12}), and next-generation wireless systems will likely employ many more antennas at base stations. 
For example, discussions to include 64-antenna base stations have already been initiated in 3GPP standardization~\cite{3gpp-64-antennas}, and ``massive'' antenna arrays with hundreds to thousands of antennas have also been proposed \cite{Marzetta2010NoCooperativeMassiveMIMO, Marzetta14MassiveMIMOOverview, Rusek12MassiveMimoOverview}.
\revision{Large arrays offer many benefits. For example, inter-beam interference can be managed with simple linear precoders/equalizers and inter-cell interference can be mitigated without requiring coordination between base stations\cite{Marzetta2010NoCooperativeMassiveMIMO, Marzetta14MassiveMIMOOverview, Rusek12MassiveMimoOverview}.
Although adding more antennas incurs the cost of more RF chains, each RF chain can be built with lower-cost components. Therefore, each RF chain carries only a fraction of the power \cite{Marzetta14MassiveMIMOOverview} as would be carried in a single-antenna base station.}

As the number of base-station antennas increases, an important question is how to enable full-duplex with a large number of antennas. 
Full-duplex 
muti-user MIMO (MU-MIMO) communication would enable the base station to transmit to multiple downlink users and receive from multiple uplink users, all at the same time and in the same frequency band, as shown in Figure~\ref{fig:problem}.
Full-duplex with many antennas presents both challenges and opportunities. 
The complexity of analog self-interference cancellation circuity grows in proportion to the number of antennas (which could potentially deter its adoption due to increased cost and complexity). At the same time, \bigArray{} full-duplex also presents an opportunity: having many more antennas than users served means that more spatial resources become available for transmit beamforming to reduce self-interference. 
 \ashuOldNote{I can't see how high-dimensionality increases complexity, it's simply the number that drives analog complexity, right?  How about "Full-duplex in \bigArray{} increases the complexity of analog in proportion to the number of antennas, potentially delaying its adoption in practice." And then jump to opportunity} \evanOldNote{Done.}

\ashuOldNote{Let's not propose removing the stage, but instead asking "what can be done without analog?" and sell it as a potential way to deploy full-duplex mode on existing radio architectures as a new mode.} \evanOldNote{I've tried to do this below}
In this work, we investigate the possibility of many-antenna full-duplex operation with current radio hardware that can either send or receive on the same band but not both, i.e. \emph{TDD\footnote{We consider only TDD radios, because FDD radios, by design, do not transmit and receive in the same band and hence cannot be transformed into in-band full-duplex.} radios without analog cancellation}.  
We propose an all-digital\footnote{\revision{By ``all-digital'' we mean that the only modifications to standard half-duplex TDD systems are in the digital domain: no additional analog components are required beyond what is already present in the half-duplex radios.}} approach called \emph{\softNull}, to enable \bigArray{} full-duplex with only digital-domain modifications. 
In the \softNull\ design, the array is partitioned into a set of transmit antennas and a set of receive antennas, and self-interference from the transmit antennas to the receive antennas is reduced by transmit beamforming. We envision that one method of using \softNull\  will be a layer \emph{below} physical layer, tasked to only reduce self-interference, and is agnostic to the upper layer processing. Thus \softNull\  can operate on the output of algorithms for downlink MU-MIMO (such as zero-forcing beamforming) without modifying their operation.
%

Transmit beamforming to null self-interference has been considered previously \cite{Day12FDMIMO,Day12FDRelay,Ashu12FDReview,Bliss07SimultTX_RX, Riihonen11FDMIMO,riihonen09, Yin13FullDuplex, Ngo14MassiveMIMORelaying, Li2015SmallCellMassiveFD, Zhang15FDMimoCapacity}, but to our knowledge, no prior work has included an experiment-based evaluation of \bigArray{} beamforming for full-duplex.  
The key departure in \softNull\  design is that \emph{our aim is not necessarily to null self-interference perfectly at each receive antenna}. Every null requires using one effective transmit antenna dimension. For a many-antenna system, self-interference is full rank and hence a nulling based self-interference scheme may end up using all available transmit degrees of freedom, leaving negligible degrees of freedom for actual downlink data transmission. Instead, our aim is to reduce self-interference  to avoid saturating the analog-to-digital conversion in the receive radio chain.  
The \softNull\ precoder minimizes the total self-interference power, given a constraint on how many \txDims\ must be preserved, where ``\txDims'' are the number of dimensions available to the physical layer for downlink communication. 
We find that the precoder to minimize total self-interference has a simple and intuitive form: the precoder is a projection onto the singular vectors of the self-interference channel corresponding to the $\userDim$ smallest singular values.

\textbf{Contribution:} Our contribution is an experiment-driven evaluation of the \revision{digital-controlled} \softNull{}-based full-duplex system using 3D self-interference channel measurements from a variety of propagation environments.  
The goal of the evaluation is to understand the conditions under which the \softNull{} system outperforms a traditional half-duplex system, and quantify how close we can approach the performance of an ideal full duplex system. 
We collect channel measurements using  a 72-element two-dimensional planar antenna array, with mobile nodes placed in many different locations, measuring self-interference channels and uplink/downlink channels both outdoors, indoors and in an anechoic chamber. The platform operates in the 2.4~GHz ISM band, with 20~MHz bandwidth. 
We use these real over-the-air channel measurements to simulate \softNull\ and evaluate its performance extensively. 
The essence of the experimental results can be captured by the following two measurement-based conclusions.  

\textbf{Self-interference reduction:}
\softNull\ enables a large reduction in self-interference while sacrificing relatively few \txDims{}. However, the amount of reduction depends on the environment: more scattering results in less suppression. 
In an outdoor low-scattering environment \softNull{} provides sufficient self-interference reduction while sacrificing only a few  \txDims{}. For example in the case of a 72-element array partitioned as 36 transmit antennas and 36 receive antennas, 50~dB of pre-analog self-interference reduction is achieved while sacrificing only 12 of the 36 available transit dimensions.  Self-interference reduction via beamforming becomes more challenging in indoor environments due to backscattering. Since backscattering makes the self-interference channel less correlated, more \txDims\ must be used to achieve the same reduction. With the same base station indoors, 20 of the 36 \txDims\ need to be used to achieve 50 dB reduction

\textbf{Data rate gains over half duplex:}
\softNull\ can provide significant rate gains over half-duplex for small cells in the case when the number of transmit antennas is much larger than the number of users.
The larger the path loss, the more challenging full-duplex operation becomes, because more self-interference reduction is required to suppress the self-interference to a power level commensurate to the power of the received uplink signal. For \softNull, more path loss means more \txDims\ must be used to suppress the self-interference to a level commensurate to the uplink signal power.
Similarly, as the number of simultaneous users served increases, the cost of using \txDims\ for self-interference reduction becomes more pronounced: not only is downlink power gain sacrificed, but downlink multiplexing gain is also sacrificed. 
For example, in the 72-antenna scenario mentioned above, with 12 users at 100~dB path loss,  the data rate achieved by \softNull{} is $12\%$ \emph{less} than half duplex, but for 4 users at 85~dB path loss the data rate improvement of \softNull{} over half duplex more than $40\%$. 
For example, in the 72-antenna scenario mentioned above, with  4 users at 85~dB path loss the data rate improvement of \softNull{} over half duplex more than $40\%$, but with
12 users at 100~dB path loss,  the data rate achieved by \softNull{} is $40\%$ \emph{less} than half duplex.
We note however, that the trend in wireless deployments is moving towards smaller cells \cite{Andrews08FemptoCellSurvey} (i.e. lower path loss) and towards operating in the regime where the number of antennas is much more than number of users served \cite{Marzetta2010NoCooperativeMassiveMIMO, Marzetta14MassiveMIMOOverview, Rusek12MassiveMimoOverview}, therefore we foresee a large application space for \softNull. 

The rest of the paper is organized as follows. Section~\ref{sec:background} describes the multi-user MIMO scenario under consideration and defines key variables and terms. Section~\ref{sec:precoder} describes the design of the \softNull\ system, in particular the self-interference suppression precoder, and gives a brief simulation example. Section~\ref{sec:setup} describes the measurement setup. Section~\ref{sec:eval} presents the results of the measurement-driven performance evaluation. Concluding remarks are given in Section~\ref{sec:conclusion}.

\section{System Definition}
\label{sec:background}
\evanOldNote{In the second draft, I have tried to pare down this section to what is minimally needed. If you feel something is missing let me know.}

We consider the multi-user system pictured in Figure~\ref{fig:problem}. 
A base station is communicating with $\numUplink$ uplink users and $\numDownlink$ downlink users. 
The base station is equipped with $\numTotal$ antennas. 
We assume the base station uses traditional radios, that is each of the $M$ antennas can both transmit and receive, but a given antenna cannot both transmit and receive at the same time. 
Therefore in full-duplex operation, $\numTx$ of the antennas transmit while $\numRx$ antennas receive, with the requirement that $\numTx + \numRx \leq \numTotal$.
Note that choice of which antennas transmit and receive can be adaptively chosen by the scheduler, but study of such adaptation is left to future work. 
In half-duplex mode all antennas are used for either transmission or reception, that is $\numTx = \numRx = \numTotal$.
The vector of symbols transmitted by the base station is $\bsTx \in \mathbb{C}^{\numTx}$, and the vector of symbols transmitted by the users is $\usrTx \in \mathbb{C}^{\numUplink}$. 
%

The signal received at the base station is 
\begin{equation}
\label{eq:uplinkChannel}
\bsRx = \upChannel \usrTx +  \selfChannel  \bsTx  + \bsNoise,
\end{equation}
where $\upChannel \in \mathbb{C}^{\numRx\times\numUplink}$  is the uplink channel matrix, $\selfChannel \in \mathbb{C}^{ \numRx \times \numTx}$ is the self-interference channel matrix, and $\bsNoise \in \mathbb{C}^{\numRx}$ is the noise at the base station's receiver. The signal received by the $\numDownlink$ downlink users is
\begin{equation}
\label{eq:downlinkChannel}
\usrRx = \downChannel \bsTx  + \interuserChannel \usrTx +  \usrNoise,
\end{equation}
where $\downChannel \in \mathbb{C}^{\numDownlink \times \numTx}$,  is the downlink channel matrix, $\interuserChannel \in \mathbb{C}^{\numDownlink \times \numUplink}$ is the matrix of channel coefficients from the uplink to the downlink users,  and $\usrNoise \in \mathbb{C}^{\numDownlink}$ is the noise at the receiver of each user.  \ashuOldNote{I am not a big fan of 's in technical document, possible to remove them?} \evanOldNote{Done.}

\begin{figure*}[ht]
\begin{minipage}{.45\textwidth}
\centering
\hspace{-75pt}
\tikzstyle{txAntenna} = [draw,fill=KeynoteBlue, inner sep=0, minimum size=10, shape=circle]
\tikzstyle{rxAntenna} = [draw,fill=red!50, inner sep=0, minimum size=10, shape=circle]
\tikzstyle{uplink} = [draw,fill=black,minimum size=1em, shape=circle]
\tikzstyle{downlink} = [draw,fill=white,minimum size=1em, shape=circle]
\tikzstyle{dummy} = [fill=white,minimum size=1em, shape=circle]
\tikz[scale=0.39, decoration={
  markings,
  mark=at position 1.0 with {\arrow{>}}}
]{
\path (-4,-2) node (arraySW) {};
\path (5,-2) node (arraySE) {};
\path (-2,-1) node (txPoint) {};
\path (3,-1) node (rxPoint) {};
\path[draw] (-4,-2) -- (5, -2) -- (5, 3) -- (-4,3) -- (-4,-2) +(-.5,2) node[anchor = east, text width=60pt] {};
\foreach \x in {-3, -2, ..., 0}
	\foreach \y in {-1, 0, ..., 2}
        		\node[txAntenna] at (\x,\y) {};
\foreach \x in {1, 2, ..., 4}
	\foreach \y in {-1, 0, ..., 2}
        		\node[rxAntenna] at (\x,\y) {};	
\path (-6,-7) node[downlink](D1) {};
\path (-7,-6) node[downlink] (D2) {};
\path (8,-7) node[uplink] (U1) {};
\path (9,-6) node[uplink] (U2) {};	
\draw[<-] (D1) to (arraySW);
\draw[<-] (D2) to node[draw, black, fill=white]{$\downChannel$} (arraySW);
\draw[->] (U1) to (arraySE);
\draw[->] (U2) to node[draw, black, fill=white]{$\upChannel$} (arraySE);
\draw[->, ultra thick, red, bend left=-130] (txPoint) to node[draw, black, fill=white] {$\selfChannel$}(rxPoint);
\draw[->, dashed, red] (U1) to node[draw, black, solid, fill=white] {$\interuserChannel$} (D1);
}
\vspace{15pt}
\caption{Multi-user full-duplex system
\label{fig:problem}}
\end{minipage}
\begin{minipage}{.55\textwidth}
\centering
\hspace{-55pt}
\scalebox{0.9}{
\tikzstyle{txAntenna} = [draw,fill=KeynoteBlue, inner sep=0, minimum size=10, shape=circle]
\tikzstyle{rxAntenna} = [draw,fill=red!50, inner sep=0, minimum size=10, shape=circle]
\tikzstyle{antenna} = [draw,fill=white,minimum size=1em, shape=circle]
\tikz[decoration={
  markings,
  mark=at position 1.0 with {\arrow{>}}}
]{
\path (-2,2.2) node (downlink)  [draw, text width=3.2cm, align=center] {MU-MIMO Downlink, $\downPrecoder$};
\path (-2,0) node (precoder)  [draw, text width=3.2cm, align=center] {\softNull\ Precoder, 
$\selfPrecoder$};
\path (1.7,2.2) node (uplink)  [draw, text width=3.2cm, align=center] {MU-MIMO \\ Uplink };
\path (1.7,0) node (cx)  [draw, text width=3.2cm, align=center] {Digital \\ Cancellation};
\foreach \x in {-.5, 0, .5}
{
		\path (precoder.north)+(\x,-.1) node (input) {};
	        	\draw[->, thick] (downlink.south)+(\x,0) to (input);
}
\foreach \x in {-1,-.5, ..., 1}
{
		\path (precoder.south)+(\x,-1) node (antenna) [txAntenna] {};
	        	\draw[->, thick] (precoder.south)+(\x,0) to (antenna);
}
\foreach \x in {-.75,-.25, ..., .75}{
		\path (cx.south)+(\x,-1) node (antenna) [rxAntenna] {};
	        	\draw[<-, thick] (cx.south)+(\x,0) to (antenna);
		\path (cx.north)+(\x,-.1) node (output) {};
	        	\draw[<-, thick] (uplink.south)+(\x,0) to (output);
}
\path (precoder.south)+(-1.5, -.5) node {$\numTx$};
\path (precoder.north)+(-1.0, .25) node {$\userDim$};
\path (cx.south)+(-1.2, -.5) node {$\numRx$};
%
\path (precoder.south)+(0,-1.25) node (txPoint) {};
\path (cx.south)+(0,-1.25) node (rxPoint) {};
\draw[->, ultra thick, red, bend right=40] (txPoint) to node[draw, black, fill=white] {$\selfChannel$}(rxPoint);
%
\path (precoder.north west)+(-.5,.6) node (leftLayerPoint) {};
\path (cx.north east)+(2,.6) node (rightLayerPoint) {};
\draw[dashed] (leftLayerPoint) to (rightLayerPoint);
\path (uplink.east)+(.2,-.1) node [anchor=west, text width=1.5cm, align=left] {\small standard\\ MU-MIMO };
\path (cx.east)+(.2,.1) node [anchor=west, text width=1.5cm, align=left] {\small \softNullLayer};
}
}
\caption{SoftNull design. First stage is \mimoLayer. Second stage is \softNullLayer, with two components: \softNull\ transmit precoder to reduce the self-interference, and receiver-side digital canceler to reduce residual self-interference. 
\label{fig:arch}}
\end{minipage}
\end{figure*}

\revision{Ongoing research is developing schedulers to select users such that the interference caused by uplink users on downlink reception is minimized~\cite{Tang15ADuplex, Kim13Janus, Singh2011FDMAC, Gai14FDScheduling, Suhas15FDScheduling} (and references within). Thus, we will make a simplifying assumption that $\interuserChannel = 0$, allowing us to focus on self-interference; we acknowledge that future work should also characterize the role of inter-node interference on overall network rate.}
In half-duplex operation the above equations are simplified: the self-interference term is eliminated in (\ref{eq:uplinkChannel}), and $\upChannel$ is a $\numTotal \times \numUplink$ matrix and $\numDownlink$ is a  $\numDownlink \times \numTotal$ matrix. 
The signaling challenge unique to full-duplex operations is how to design $\bsTx$ such that the self-interference is small, while still providing a high signal-to-interference-plus-noise ratio (SINR) to the downlink users.  \revision{One of the major challenges in many-antenna MIMO systems is the acquisition of downlink channel state information (CSI) at the base station, such that $\downChannel$   can be used to compute the downlink precoder. 
The need to acquire CSI is universal to all many-antenna MIMO beamforming systems, and  not specific to full-duplex implementation.  A large body of work is devoted to studying methods for obtaining channel state information with low overhead, such as \cite{Marzetta2010NoCooperativeMassiveMIMO, Shepard:2012:APM:2348543.2348553, Marzetta13Pilots, Heath13ChannelAging, Marzetta11PilotContam}, and references therein. 
}


\section{\softNull~Design}
\label{sec:precoder}


\ashuOldNote{The modularity of \softNull\ is not emphasized in the introduction.} \evanOldNote{I added it.}
The physical layer design for \softNull\ is depicted in Figure~\ref{fig:arch}. 
We propose a two-stage approach. The first stage is \mimoLayer\ for which conventional precoding and equalization algorithms can be used. The second stage is the  \softNullLayer\ stage, which reduces self-interference via transmit beamforming and digital self-interference cancellation. The advantage of this two-stage approach is that \softNull\ can be incorporated as a modular addition to existing MU-MIMO systems. 
The disadvantage is that the performance may be sub-optimal due to the two-stage constraint.  
Joint precoder design for MU-MIMO downlink and self-interference reduction is a topic for future work but is outside the scope of this paper. \lin{layer is a loaded word. You now have a two-layered design within the physical layer. Perhaps ``stage''? Also later you say SoftNull has two components. From Figure 2, it has four components. The self-interference reduction ``layer'' has the two components you talked about.} \evanNote{Good points. I have resolved this per your recommendation.}
\clay{Perhaps justify this `sub-optimal' design by mentioning that it is computationally feasible/practical in real systems?} \evanNote{I feel the justification is already there in the statement `` can be incorporated as a modular addition to existing MU-MIMO systems.'' Beyond that, we have not done any analysis that can support a claim of practicality or computational feasibility. I am up for ideas if you have any, but I think it's fine as is.}

The \softNullLayer\ stage of \softNull\ has two components: a transmitter-side precoder to reduce self-interference and a receiver-side digital canceler to reduce residual self-interference. 
Digital cancellation is  well understood, and we believe existing techniques are sufficient for practical use; see e.g.~\cite{Duarte2012FullDuplexWiFi,Katti2014FullDuplexMIMO}. 
Thus, in this section, we focus on the  design of the \softNull\ precoder. 
We assume that the decision on the partitioning of the transmit and receive antennas $(\numTx,\numRx)$ is made by a higher layer operation, based on the network needs. 


\subsection{Precoder Design}
As shown in Figure~\ref{fig:arch}, the downlink precoder has two stages, a standard MU-MIMO downlink precoder, $\downPrecoder$, followed by the \softNull~precoder, $\selfPrecoder$.  
The goal of the  \softNull~precoder, $\selfPrecoder$, is to suppress self-interference.  The goal of the downlink precoder, $\downPrecoder$, is for the signal received by each user to contain mostly the signal intended for that user, and little signal intended for other users. 
The standard MU-MIMO downlink precoder, $\downPrecoder$, controls $\userDim$ effective antennas.  The \softNull\ precoder maps the  signal on the $\userDim$ effective antennas to the signal transmitted on the $\numTx$ physical transmit antennas, as shown in Figure~\ref{fig:arch}.
Let $\downSyms \in \mathbb{C}^{\numDownlink}$ denote the vector of symbols that the base station wishes to communicate to each of the $\numDownlink$ downlink users. 
We constrain both stages to be linear, such that $\downPrecoder$ is a $\userDim \times \numDownlink$ complex-valued matrix and  $\selfPrecoder$ is a $ \numTx \times \userDim$ matrix. The signal transmitted on the base station antennas is then $\bsTx = \selfPrecoder \downPrecoder \downSyms $. 
\clay{I modified this previous sentence because the downlink precoder doesn't (and can't) ensure each user `only' receives their intended signal. (I first said maximize, but that technically isn't correct either.)} \evanNote{But increase relative to what?  I have rewtritten again to make a happy medium.}

\subsubsection{Standard MU-MIMO downlink precoder}

The standard MU-MIMO downlink precoder, $\downPrecoder$, does not need to have knowledge of both the self-interference channel and the downlink channel. Rather the downlink precoder, $\downPrecoder$,  only needs to know the \emph{effective} downlink channel, $\effDownChannel = \downChannel\selfPrecoder$, that is created by the \softNull\ precoder operating on the physical downlink channel. Note that $\effDownChannel$ can be estimated directly by transmitting/receiving pilots along the $\userDim$ \txDims. 
For the standard MU-MIMO downlink precoder, standard algorithms such as zero-forcing beamforming or matched filtering can be used. For example, in the case of zero-forcing beamforming, the MU-MIMO downlink precoder, $\downPrecoder$,  is the Moore-Penrose (right) pseudoinverse of the effective downlink channel:
\begin{equation}
\downPrecoder = \downPrecoder^\mathsf{(ZFBF)}  \equiv  \alpha^\mathsf{(ZFBF)} \effDownChannel(\herm{\effDownChannel}\effDownChannel)^{-1}, 
\end{equation}
where $\alpha^\mathsf{(ZFBF)}$ is a power constraint coefficient.

\label{sec:precoderDesign}

\subsubsection{\softNull~precoder}

The goal of the \softNull\ precoder is to reduce self-interference while preserving a required number of \txDims, $\userDim$, for the standard MU-MIMO downlink transmission. As shown in Figure~\ref{fig:arch} the \softNull\ precoder has $\userDim$ inputs as effective antennas, and $\numTx$ outputs to the physical antennas. 
We assume that the \softNull\ precoder has knowledge of the self-interference channel, $\selfChannel$. 
Our goal is to minimize the \emph{total} self-interference power while maintaining $\userDim$ \txDims.  
Our choice to minimize total self-interference, rather than choosing a per-antenna metric is twofold: 
(i) Minimizing total self-interference gives the precoder more freedom in its design. Instead of creating nulls to reduce self-interference at specific antennas, it can optimize placement of nulls such that each null can reduce self-interference to multiple receive antennas. 
(ii) As is shown in the following, minimizing the total self-interference power leads to a closed-form solution.  
We therefore formulate the precoder design problem as:
\begin{align}
\label{eq:optimStatement}
	\selfPrecoder =& \underset{\precoder}{\argmin} \|\selfChannel \precoder \|_F^2  \\ \nonumber
	&\text{subject to } {\herm{\precoder}\precoder = \identity{\userDim}{\userDim}}.
\end{align}
%
The squared Frobenius norm, $\|\cdot\|_F^2$, measures the total self-interference power.
The constraint,  $\herm{\precoder}\precoder = \identity{\userDim}{\userDim}$, forces  the precoder to have $\userDim$ orthonormal columns,  and hence ensures that $\userDim$ \txDims\ are preserved for MU-MIMO downlink signaling. \ashuOldNote{I find transmit dimensions and effective antennas terminology confusing. In the early part, when explaining layered approach, say that the softnull ensures that the above layers have certain number of effective transmit dimensions and draw antenna lines as in our presentations. }\evanOldNote{I have modified to ``\txDims'' terminology everywhere, and have eliminated ``transmit dimensions''. Also have added the comment you mentioned at the top of the paragraph.}

It is shown in Appendix~\ref{sec:precoderProof} that the above optimization problem (\ref{eq:optimStatement}) has the following closed-form intuitive solution. 
The optimal self-interference precoder is constructed by projecting onto the  $\userDim$ left singular vectors of the self-interference channel corresponding to the smallest  $\userDim$ singular values. Precisely, 
\begin{equation}
\label{eq:svdPrecoder}
\selfPrecoder = \left[ \singVecRight{\numTx-\userDim+1}, \singVecRight{\numTx-\userDim+2}, \dots, \singVecRight{\numTx}\right], 
\end{equation}
where $\selfChannel = \singLeft \singVals \herm{\singRight}$ is the singular value decomposition of the self-interference channel
($\singLeft$ and $\singRight$ are unitary matrices and $\singVals$ is a nonnegative diagonal matrix whose diagonal elements are the ordered singular values)
and $\singVecRight{i}$ is the $i$th column of $\singRight$. 
Essentially, the \softNull\ precoder is finding the  $\userDim$-dimensional subspace of the original transmit space, $\mathbb{C}^{\numTx}$, which presents the least amount of self-interference to the receiver. 
\subsection{\softNull~Simulation~Example}
\label{sec:example}

\revision{
To help clarify the \softNull{} design, we provide a simple simulation example that illustrates how \softNull\ reduces self-interference by sacrificing \txDims{}.
Figure~\ref{fig:simArray} shows a $3\times6$ $(M=18)$ planar array that is the basis of the simulation. The space between adjacent antenna is half a wavelength. We consider an even $(\numTx,\numRx) =(9,9)$, partition of transmit and receive antennas,  as shown in Figure~\ref{fig:simArray}, where blue circles on the left correspond to the  $3\times3$ transmit subarray, and the red circles on the right correspond to the $3\times3$ receive subarray.

Fields are computed using MATLAB's full-wave method-of-moments electromagnetic solver to compute the electric field distribution produced by the transmit antennas in the vicinity of the receive antennas. The full-wave solver enables near-field effects as well as the impact of mutual coupling among the transmit antennas to be captured in the simulation. The antenna's used are vertically polarized patch antennas, tuned to 2.4~GHz operation. The patch's resonant dimension is $L = 48.7$~mm,  and an air dielectric between patch and ground is assumed. 


\begin{figure*}[h!]
\centering
\subfigure[Simulated array partition: the left blue circles denote transmit antennas and right red circles denote receive antennas.\label{fig:simArray}]{\centering
\hspace{1em}\input{DocGraphics/simArray.tex}\hspace{1em}
}\hspace{1em}
\centering
\subfigure[Radiation pattern (for beam steered to broadside of array) for different values of number of \txDims, $\userDim$. 
\label{fig:visualizeGain}]{\centering
\includegraphics[width=0.45\textwidth, clip=true, trim = 3cm 6cm 3cm 0]{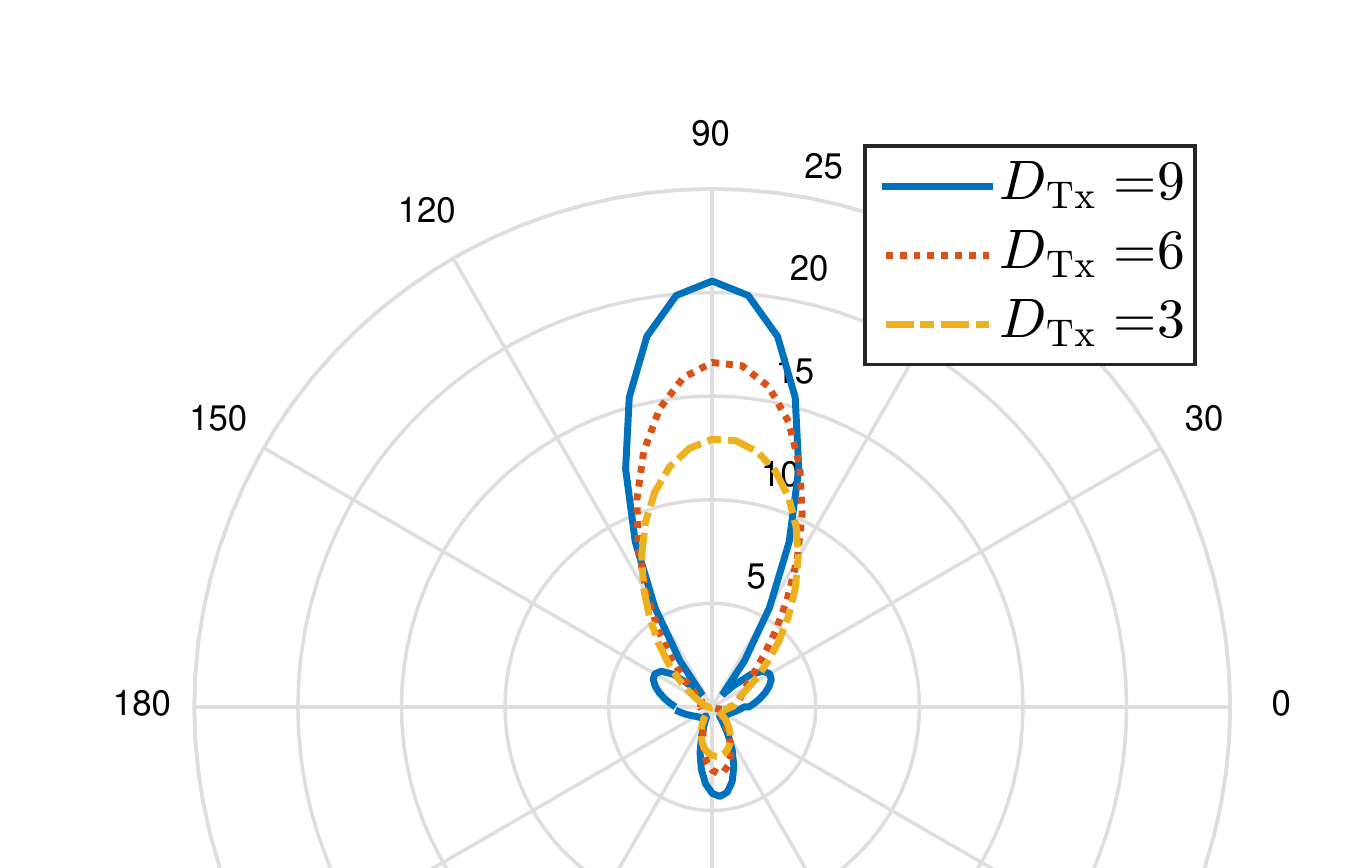}
}
\centering
\subfigure[Distribution of field strength for different values of number of \txDims, $\userDim$. 
\label{fig:visualizeFields}]{ \centering
\includegraphics[width=\textwidth, clip= true, trim = 4cm 0  4cm 0]
{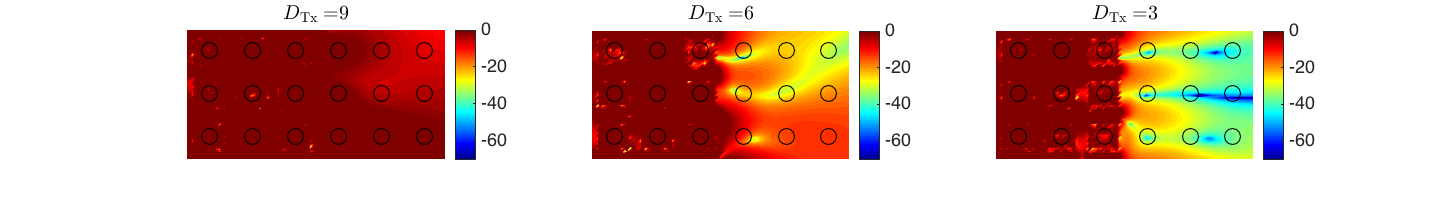}
}
\caption{\revision{Simulation example of \softNull{} precoder operation. (a) The array simulated. (b) Gain of radiation pattern steered to broadside.  (c) Distribution of field strength around the receive antennas. \label{fig:exampleSim}}} 
\end{figure*}

Figure~\ref{fig:visualizeFields} shows the radiated field distribution, in the vicinity of the received antennas, as a function of the number of \txDims, $\userDim.$  
First consider the case where $\userDim=9=\numTx$, in which no \txDims{} are given up for the sake of self-interference reduction: all the receive antennas receive very high self-interference. 
Then, in the case where  $\userDim=6$, and three effective antennas are given up for self-interference reduction, the \softNull\ precoder essentially steers ``soft'' nulls  towards the receive potion of the  array, achieving a small amount of self-interference reduction. 
In the case of $\userDim=3$, the six \txDims{} sacrificed allow the \softNull\ precoder  to significantly reduce self-interference at each receive antenna. The overall trend is that as more \txDims\ are given up for the sake of self-interference reduction, the more freedom \softNull\ has to create a radiated field pattern which reduces self-interference. 
Due to computation limits, we have considered a small array in this simulation, for which the self-interference reduction is modest. However, in the large-array experiments that follow, we will see that having a large array enables more design freedom, and thus better self-interference reduction. 
Figure~\ref{fig:visualizeGain} illustrates the downside to using \txDims\ for self-interference suppression: reduced transmit gain. In Figure~\ref{fig:visualizeGain} we plot the radiation pattern (i.e., far-field power gain) for a beam steered towared the array's broadside. 
As we give up more \txDims{} for the sake of self-interference reduction, the array gain is decreased. Therefore in the following sections, we will carefully evaluate \emph{whether the benefit in self-interference reduction is worth the loss in beamforming gain.}
}



\section{Channel Measurement Setup}
\label{sec:setup}
\label{sec:campaign}

\begin{figure*}[htbp]
\centering
\subfigure[Planar antenna array interfaced to WARP radios.\label{fig:array}]{
	\centering
	\hspace{0em}
	\includegraphics[width = 0.3\textwidth, clip=true, trim = 0 5cm 0 20cm]{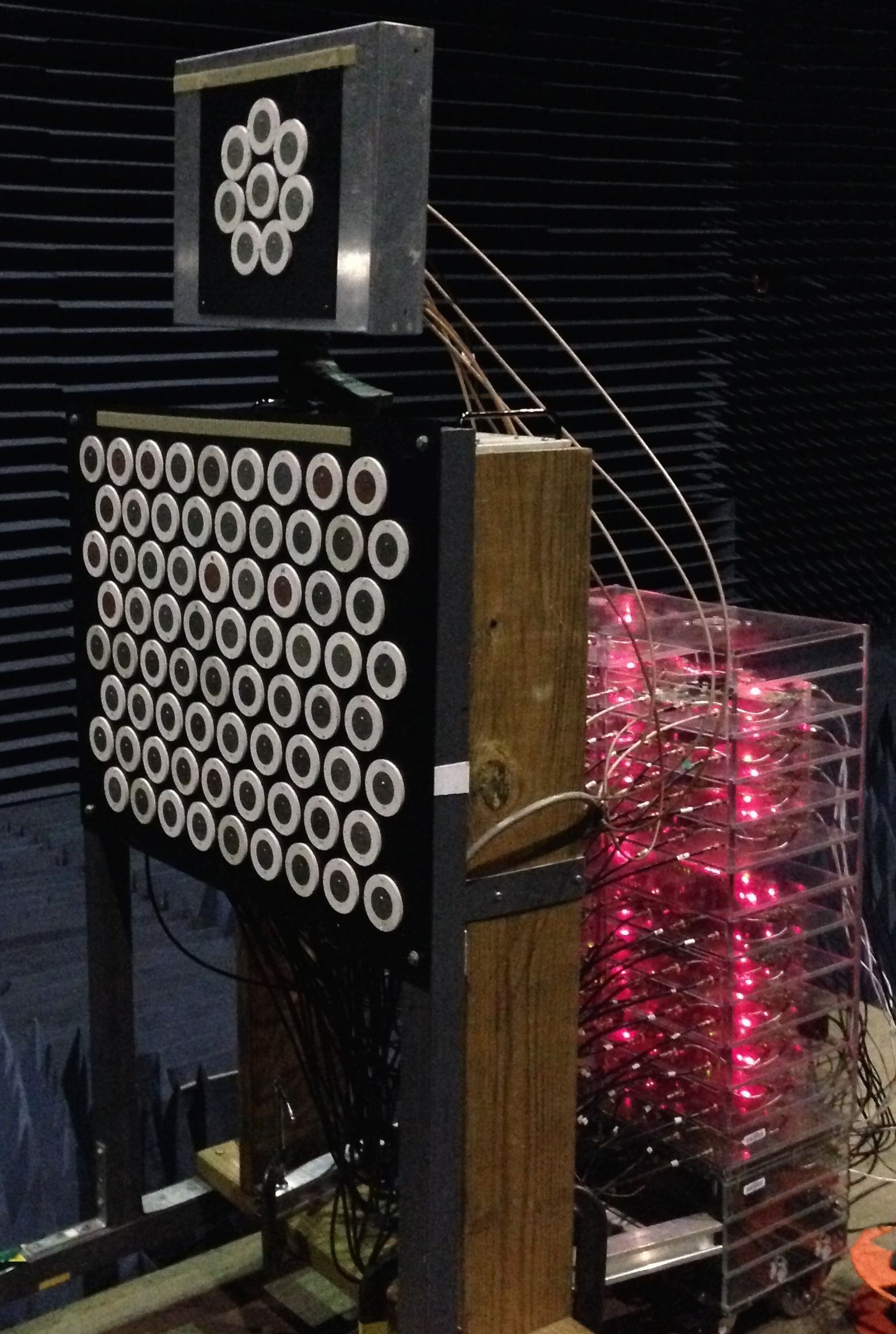}	 
	\hspace{1em}
	}
	\subfigure[72-element planar array.\label{fig:arrayPic}]{
		\includegraphics[width = 0.45\textwidth, clip=true, trim = 0 0 0 6cm]{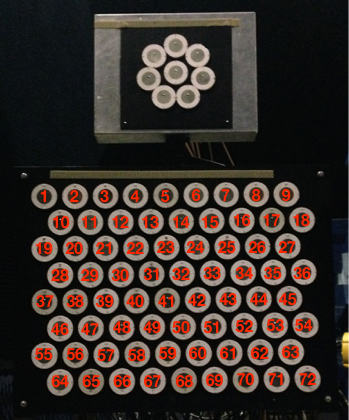}	
	}
	\caption{Platform for channel measurements \label{fig:hardware}}
\end{figure*}


To evaluate the performance of \softNull, measurements of real self-interference channels and array-to-client channels were collected using the ArgosV2 measurement platform, \cite{ArgosV2}, shown in Figure~\ref{fig:array}.
The platform consists of an array of 72 2.4~GHz patch antennas interfaced to 18 WARP v3 boards~\cite{WARP}, each with 4 programmable radios.  
This platform enables 72 base station antennas (transmit or receive).
Also four mobile clients are emulated using WARPv3 radios.
The antenna array, shown in Figure~\ref{fig:arrayPic} uses custom 2.4~GHz half-wave circular patch \cite{BalanisText} antenna elements in a hexagonal grid spaced at \revisionTwo{76 mm} apart (0.6$\lambda$). The antenna elements have roughly 6 dBi gain at broadside.  
\revision{See \cite{BalanisText, bahl80CircPatch} and references therein for more details on patch antenna design and patch antenna radiation characteristics (input impedance, polarization, radiation patterns, etc.).}

\begin{figure*}[htbp]
\renewcommand{\subcapsize}{\normalsize}
\centering
	\subfigure[Anechoic chamber\label{fig:chamber}]{
	\centering
	\includegraphics[width = 0.22\textwidth]{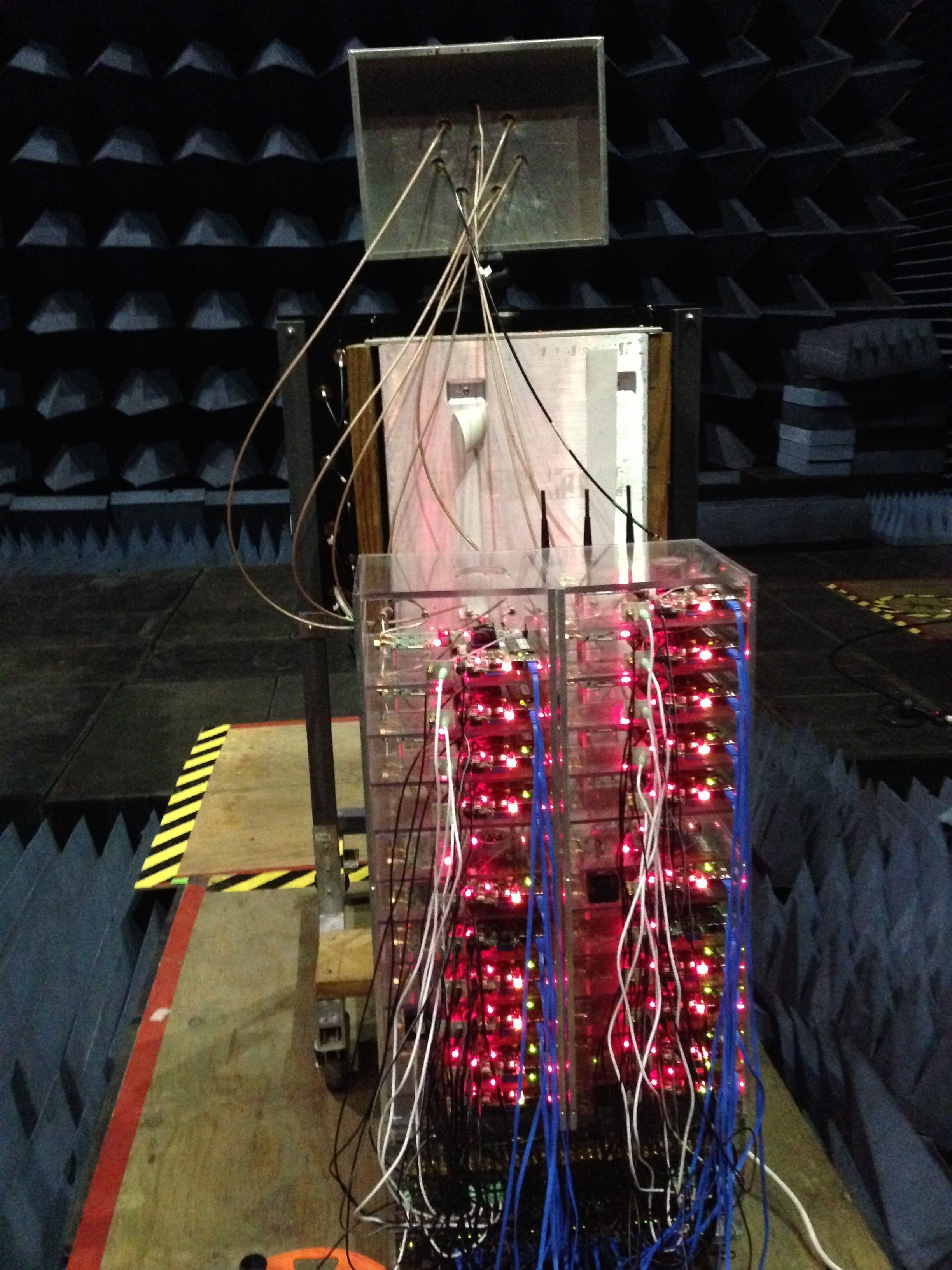}
	}
	\subfigure[Outdoor deployment \label{fig:outdoors}]{
	\centering
	\includegraphics[width = 0.3\textwidth]{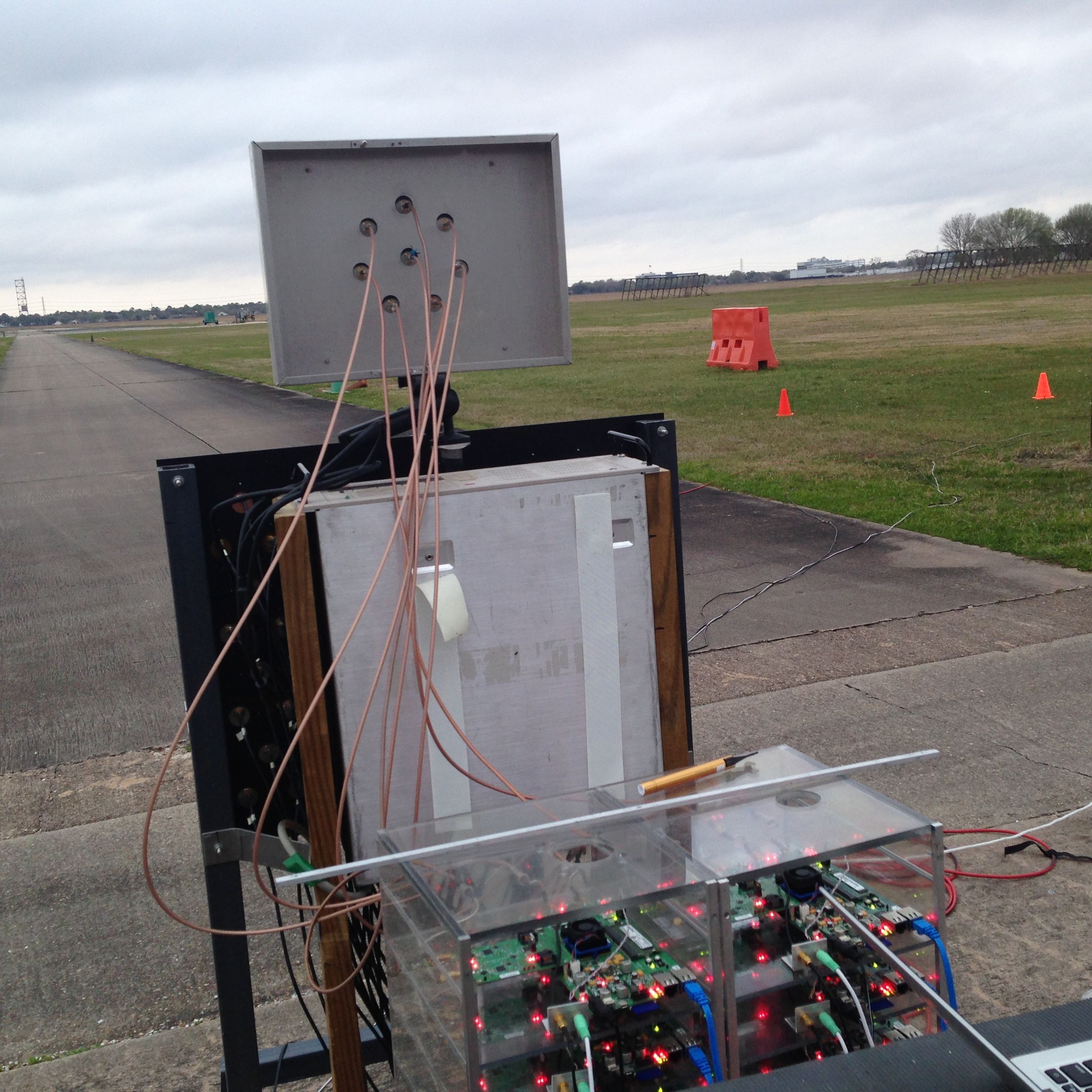}
    	}
	\subfigure[Indoor deployment \label{fig:indoors}]{
	\centering
	\includegraphics[width = 0.4\textwidth]{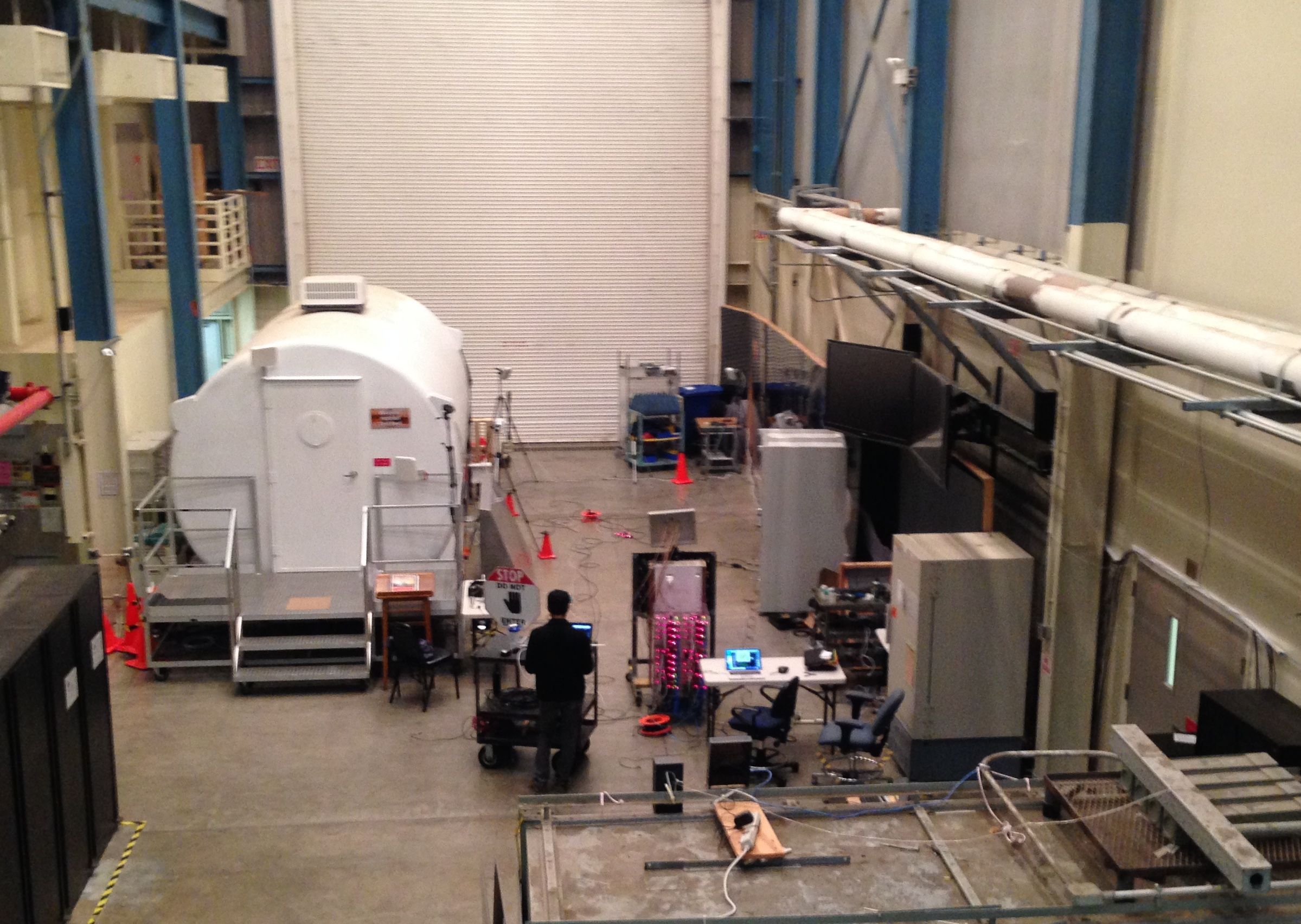}
	}
\caption{Experiment setup}
\label{fig:environs}
\end{figure*}

To collect traces of real self-interference channel for the 2D array, 20~MHz wideband channel measurements were performed in a diverse set of environments  shown in Figure~\ref{fig:environs}. 
The measured channel traces enable subsequent analysis to obtain an in-depth understanding of coupling between base station antennas, to explore Tx/Rx partitions, and to ultimately simulate real-world performance with clients. 
Measurements were taken in an anechoic chamber deployment, Figure~\ref{fig:chamber}, an outdoor deployment, Figure~\ref{fig:outdoors}, and in a highly scattered indoor deployment,  Figure~\ref{fig:indoors}. 
The outdoor deployment was in an open field, with very few obstructions to cause scattering. 
Finally the indoor deployment was in a very rich scattering environment, with metal walls and the array placed near a metallic structure as shown in Figure~\ref{fig:indoors}.

For each deployment, the four clients were placed at three different locations each and channel measurements were performed --- both for the $72\times72$ self-coupling of the array, and the $72\times4$ matrix of downlink/uplink channels for each placement. 
In all, more than 12 million wideband channels were measured, providing more \revisionTwo{than} 40 GB of channel traces for the evaluation of \softNull\ performance. 
\revision{Our channel traces are publicly available \cite{softNullData} for other researchers to leverage in their own simulations. }

\begin{figure*}[htbp]
\renewcommand{\subcapsize}{\normalsize}
\centering
	\subfigure[Outdoor \label{fig:couplingOutdoors}]{
	\centering
	\includegraphics[width = 0.47\textwidth]{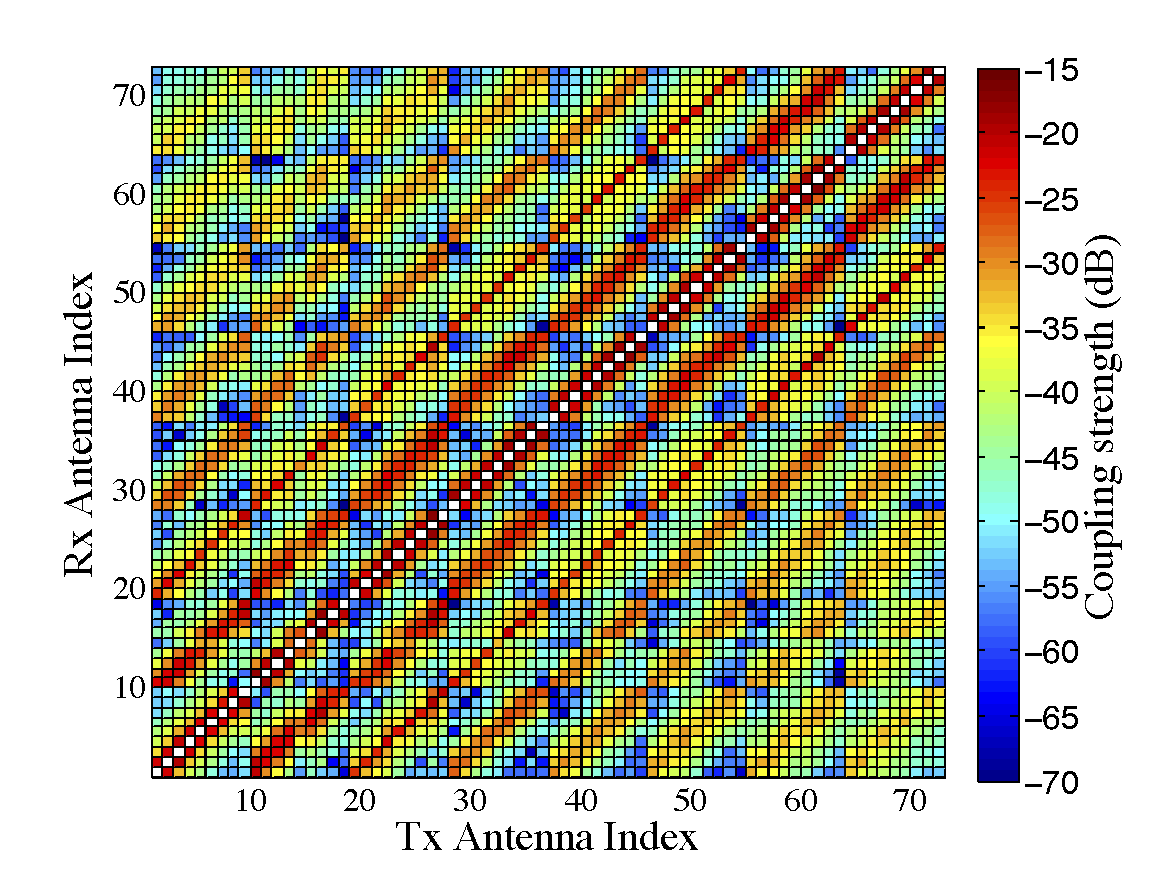}
    	}
	\subfigure[Indoor \label{fig:couplingIndoors}]{
	\centering
	\includegraphics[width = 0.47\textwidth]{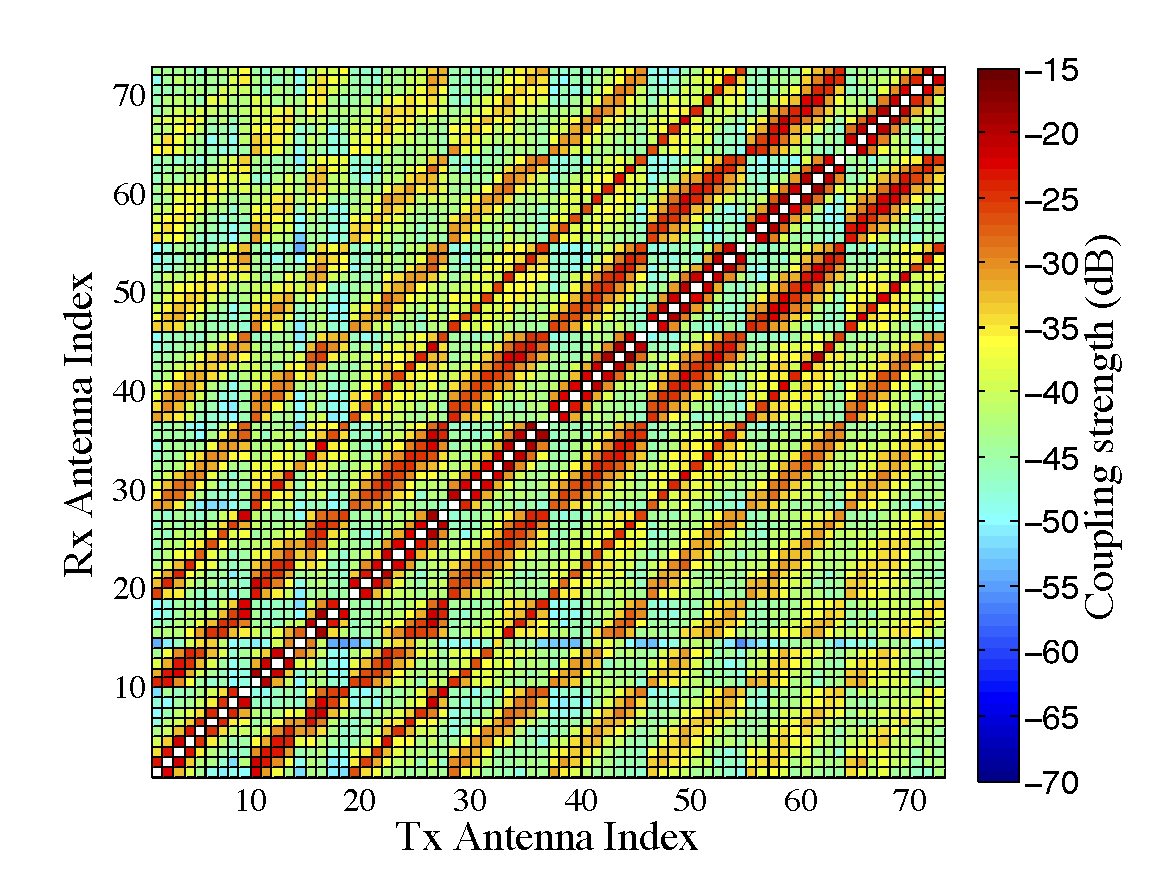}
	}
\caption{Array self-coupling}
\label{fig:coupling}
\end{figure*}

\revision{
Here we present the results on of the amount of coupling (also called ``crosstalk'' and ``mutual coupling'' in the literature) among the elements of the antenna array. Figure~\ref{fig:coupling} shows the strength of the coupling between all pairs of antennas on the array. The color of element $(i,j)$ denotes the average strength of the channel from antenna $i$  to antenna $j$, where the enumeration of the antenna elements is row-wise as shown in Figure~\ref{fig:arrayPic}. Figure~\ref{fig:couplingOutdoors} shows the average coupling among the outdoor measurements and Figure~\ref{fig:couplingIndoors} shows the average coupling indoors\footnote{For the sake of space the results for the anechoic chamber have been left off, as they were very similar to the results for outdoors and did not yield additional insight. The full datasets can be downloaded at \cite
{softNullData}.}. The banded structure in the matrix is due to counting row-wise (every 9 elements the row ends and a new row begins). 
Only the coupling between distinct antennas was measured (i.e. the hardware was not capable of measuring the reflection coefficient for individual antennas), hence the empty diagonal.  As expected, there is significant variation in the amount of coupling over the array: antennas adjacent  to one another can have a coupling as strong as $-15$~dB. In the outdoor environment, elements which are far apart can have coupling less than $-60$~dB, due to the separation and the directionality of the patch antenna elements. In the indoor environment, however, the the coupling between even far away antennas never falls much below $-50$~dB. The reason for increased coupling indoors is backscattering: for far apart antennas, the coupling due to backscattering from objects in the environment is actually stronger that the direct-path coupling. This phenomenon of backscattering becoming the self-interference bottleneck for well-isolated antennas was also reported in \cite{Everett2013PassiveSuppressionFD}. 
}

\section{\softNull\  Evaluation}
\label{sec:eval}
In this section, we utilize the collected channel traces to analyze the performance of \softNull\ in several aspects. 
First, we consider how the partition of the whole array into transmit and receive sub-arrays impact the ability of \softNull\ to reduce self-interference. 
Next, we study the impact of the propagation environment on the total self-interference reduction by \softNull. 
Finally, we consider the uplink and downlink data rate that \softNull\ can deliver to clients, and compare with half-duplex and ideal full-duplex systems.

\subsection{Antenna Array Partitioning}\label{sec:partition}
Previously, we presented the optimal precoder design of \softNull, for a given $\numRx \times \numTx$ self-interference channel. 
We now consider how to best partition the $M$ antennas into the set of $\numTx$ transmit antennas and  $\numRx$ receive antennas. 
Due to the combinatorial nature of the problem, finding the optimal antenna sets is computationally difficult. 
For example, if $M=72$ and $\numTx = 36$, then there are ${72 \choose 36}\approx 4.4\times10^{20}$ possible combinations of transmit antenna sets. 
Therefore we focus on empirical insights using the traces collected via channel measurements to evaluate and compare several heuristic choices for partitioning the array.  

\subsubsection{Heuristic Partitions Considered}
\begin{figure*}[htbp]
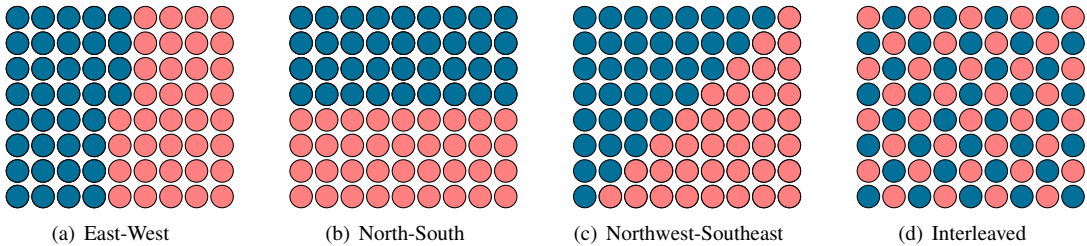

\centering
	\subfigure[East-West\label{fig:EastWestSplit}]{
	\centering
	\input{DocGraphics/36x36LeftRightSplit.tex}	
	}
	\subfigure[North-South \label{fig:NorthSouthSplit}]{
	\centering
	\input{DocGraphics/36x36NorthSouth.tex}
    	}
	\subfigure[Northwest-Southeast \label{fig:DiagSplit}]{ 
	\centering
	\input{DocGraphics/36x36NW-SE.tex}
	}
	\subfigure[Interleaved \label{fig:Interleaved}]{
	\centering
	\input{DocGraphics/36x36InterLeaved.tex}
	}
\caption{Tx/Rx partitions heuristics: blue is transmit, red is receive.}
\label{fig:partitions}
\end{figure*}

Intuitively, we recognize that \softNull\ will perform the best when the power in the self-interference channel is concentrated within a fewer number of eigenchannels.   
It has been demonstrated both analytically and experimentally \cite{TsePoon05DOF_EM,TsePoon06EmagInfoTheory, Bolcskei03MimoPropagation, Spencer2000ClusteredChannels} that as the spread of the angles-of-departure from the transmitter to the receiver is decreased, the signal received at each receive antenna becomes more correlated. More correlated signals leads to the first few eigenvalues become more dominant, which is exactly the desirable situation for \softNull. 
Contiguous linear partitions of the array (one side transmit, other receive) limit the angular spread of the angles-of-departure to/from the transmitter to the receiver, since  all the interference is coming from only one ``side'' of the array. For example, in the North-South partition of Figure~\ref{fig:NorthSouthSplit} the angular spread of angles-of-arrival is less than 180 degrees for all receive antennas, since all interference is coming from the ``North.''  
Figure~\ref{fig:partitions} shows three proposed partitions based on the above heuristic of linear contiguous partitioning to limit angular spread: East-West, North-South, and Northwest-Southwest partitions are shown in Figures~\ref{fig:EastWestSplit},~\ref{fig:NorthSouthSplit}, and \ref{fig:DiagSplit}, respectively.  
In the figure, an even split between the number of transmit and receive antennas is assumed. 

As a comparison, we also consider the interleaved partition shown in Figure~\ref{fig:Interleaved}. 
If our heuristic of minimizing angular spread is effective, then we would expect the interleaved partition to be a near worst-case partition. In the interleaved partition receive antennas will experience interference arriving at every possible angle. 
In addition to the deterministic interleaved partition, we also compare against the average measured performance of $10,000$ randomly chosen partitions.

\ashuOldNote{I suggest to use self-interference reduction versus self-reduction suppression.}
\evanOldNote{Done}
\ashuOldNote{Also why 50dB mark is relevant, we are missing the context of prior work which also got 40-50dB from analog. Cite my JSAC full-duplex review paper for calculations.}
\evanOldNote{Done}

\subsubsection{Evaluation of the Heuristic Partitions}
\begin{figure}[htbp]
\renewcommand{\subcapsize}{\normalsize}
\centering
{\small
\begin{tikzpicture}
\begin{axis}[%
scale only axis,
width=0.35\textwidth,
height=0.25\textwidth,
xtick={4,8,...,36},
xlabel={Effective antennas, $\userDim$},
ylabel={SI Reduction (dB)},
xlabel near ticks,
ylabel near ticks,
axis on top,
legend entries={
	Random,
	North-South,
	East-West,
	NW-SE,
	Interleaved
},
legend style={at={(1.2,0.7)},anchor=east,nodes=right}
]
%
\addplot [color=KeynoteGray,  line width=.75pt, mark=triangle]
	plot [error bars/.cd, y dir=both, y explicit,  
		error bar style={line width=.75pt},
	  	error mark options={rotate=90,line width=.5pt}]
	table [x = dof, 
		y = avgSup,
		]{./DocData/36tx36rxRandomChoice.dat};
\addplot [color=KeynoteRed, line width=.75pt, mark=diamond]
	plot [error bars/.cd, y dir=both, y explicit,  
		error bar style={line width=.75pt},
	  	error mark options={rotate=90,line width=.5pt}]
	table [x = dof, 
		y = avgSup,
		]{./DocData/36tx36rxTop-Down.dat};
\addplot [color=KeynoteBlue,  line width=.75pt, mark=square]
	plot [error bars/.cd, y dir=both, y explicit,  
		error bar style={line width=.75pt},
	  	error mark options={rotate=90,line width=.5pt}]
	table [x = dof, 
		y = avgSup,
		]{./DocData/36tx36rxLeft-Right.dat};
\addplot [color=KeynoteYellow,  line width=.75pt, mark=o]
	plot [error bars/.cd, y dir=both, y explicit,  
		error bar style={line width=.75pt},
	  	error mark options={rotate=90,line width=.5pt}]
	table [x = dof, 
		y = avgSup,
		]{./DocData/36tx36rxCorners.dat};
\addplot [color=KeynoteGreen,  line width=.75pt, mark=x]
	plot [error bars/.cd, y dir=both, y explicit,  
		error bar style={line width=.75pt},
	  	error mark options={rotate=90,line width=.5pt}]
	table [x = dof, 
		y = avgSup,
		]{./DocData/36tx36rxInterleaved.dat};
\end{axis}
\end{tikzpicture}
}	
\caption{Achieved tradeoff between self-interference reduction and number of \txDims{} remaining for downlink signaling, $\userDim$. Several different methods of partitioning  72-element array into and even $(\numTx, \numRx)$ = (36,36) Tx/Rx split are compared.  \label{fig:partitionsResults}}
\end{figure}
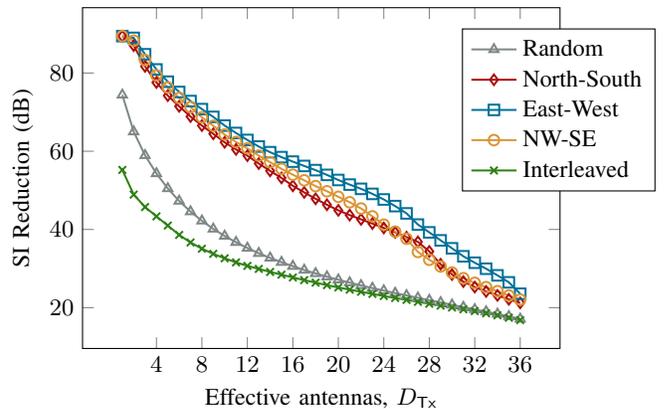

To assess the performance of these heuristics, we directly measured the self-interference channel response in an anechoic chamber using the 72-element rectangular array and a $(\numTx,\numRx) = (36,36)$ partition of transmit and receive elements.  We consider the self-interference channel measurements performed in the anechoic chamber, as this is the most repeatable scenario. Figure~\ref{fig:partitionsResults} shows  the tradeoff between self-interference (SI) reduction and number of effective antennas, $\userDim$. As $\userDim$ decreases from its maximum value of $\userDim=\numTx=36$, the amount of self-interference reduction achieved by \softNull{} improves. 
We remind the reader that $\userDim$ is the number of effective antennas preserved for downlink signalling, and thus  $(\numTx - \userDim)$ is the number of effective antennas leveraged for self-interference reduction.  As $\userDim$ decreases we are ``giving up'' effective antennas for the sake of improved self-interference reduction. Therefore as  $\userDim$ decreases, we expect to observe better self-interference reduction, as we see in the case of Figure~\ref{fig:partitionsResults}. We see in Figure~\ref{fig:partitionsResults}, that the tradeoff achieved for the contiguous partitions is much better than that achieved for the random and interleaved partitions. 

Consider the tradeoff between self-interference reduction and  \txDims\ shown in Figure~\ref{fig:partitionsResults}.
Typical analog cancellation circuits provide 40-50 dB self-interference reduction \cite{Ashu12FDReview}. Therefore, an interesting point of observation is how many \txDims\ can be preserved while achieving the $\approx50~$dB self-interference reduction similar to that of an analog canceler.  
For the random partition, we can only preserve $6$ of the maximum $36$ \txDims\ while achieving  $>50$~dB self-interference reduction, but for all of the contiguous partitions,  we can achieve $>50$~dB reduction with at least 16 \txDims\ preserved. 
The best performing partition is the East-West partition, which is in line with our heuristic: among the considered partitions, the East-West partition is the one with minimum angular spread between the transmit and receive partitions, since it splits the array along its smallest dimension (array is wider than tall). The interleaved partition performs even worse than the average of random partitions, emphasizing the importance of selecting contiguous partitions. 
Finally, note the large impact of the partition type on the tradeoff between self-interference reduction and number of effective antennas.
For $\userDim \in [3,22]$, the East-West partition enables \softNull~to achieve more than $25$~dB better self-interference reduction than the average of partitions chosen at random. 

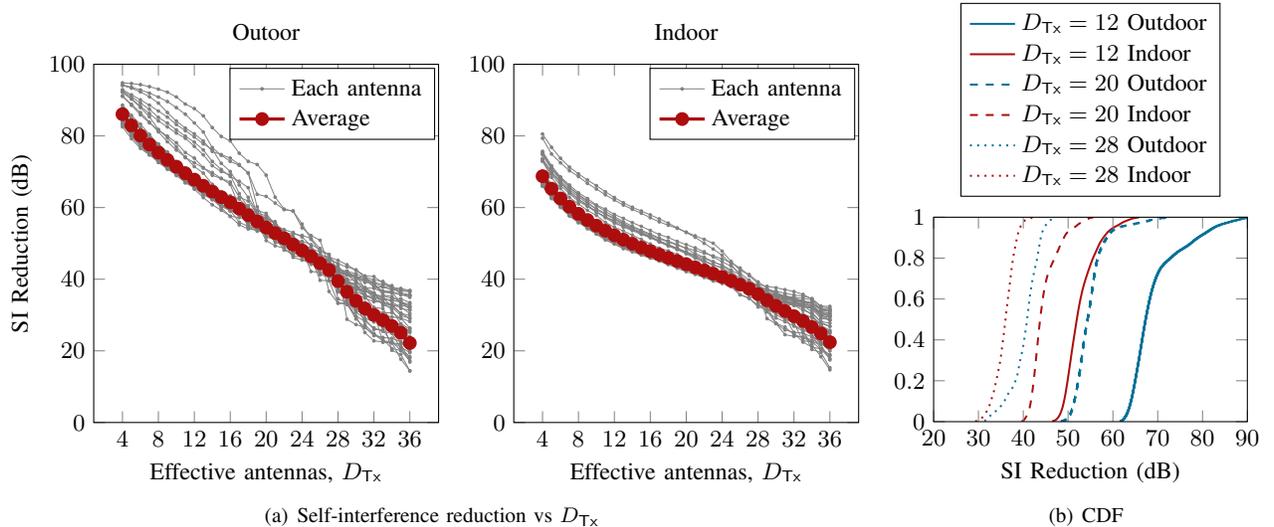
\begin{figure*}[htbp]
\centering
	\subfigure[Self-interference reduction vs $\userDim$  \label{fig:suppressionVsDofs}]{
	\centering
	{\small \begin{tikzpicture}
          \begin{groupplot}
          [ 
            group style={
            	group name=my plots,
            	group size=2 by 1,
            	xlabels at=edge bottom,
            	ylabels at=edge left,
        		},
           	xlabel={Effective antennas, $\userDim$},
            ymin = 0, ymax = 100,
            xtick={4,8,...,36},
            width = 0.34\textwidth, height=0.35\textwidth,
            legend pos = north west,
            cycle list name = color list,
            xlabel near ticks,
	    legend entries={Each antenna, Average},
          legend style={at={(0.99,0.99)},anchor=north east,nodes=right}
            ]
\nextgroupplot[ylabel={SI Reduction (dB)}, ylabel near ticks, title={Outoor}]
	\addplot +[mark=*, mark size=.5, very thin, gray] table [y index = 1] {DocData/SuppressionPerAntenna-East-West-36Tx36Rx-Outdoor-InfdBPathLoss.dat};
          \foreach \yindex in {2,...,36}
            {\addplot +[mark=*, mark size=.5, very thin, gray, forget plot] table [y index = \yindex] {DocData/SuppressionPerAntenna-East-West-36Tx36Rx-Outdoor-InfdBPathLoss.dat};
            } 
            \addplot [mark=*, very thick, draw=black, color=KeynoteRed] table {DocData/SuppressionMean-East-West-36Tx36Rx-Outdoor-InfdBPathLoss.dat};
\nextgroupplot[title={Indoor}]
	\addplot +[mark=*, mark size=.5, very thin, gray] table [y index = 1] {DocData/SuppressionPerAntenna-East-West-36Tx36Rx-Indoor-InfdBPathLoss.dat};
          \foreach \yindex in {2,...,36}
            {\addplot +[mark=*, mark size=.5, very thin, gray, forget plot] table [y index = \yindex] {DocData/SuppressionPerAntenna-East-West-36Tx36Rx-Indoor-InfdBPathLoss.dat};
            } 
            \addplot [mark=*, very thick, draw=black, color=KeynoteRed] table {DocData/SuppressionMean-East-West-36Tx36Rx-Indoor-InfdBPathLoss.dat};
\end{groupplot}
\end{tikzpicture}}
	}
	\subfigure[CDF 
\label{fig:suppressionCDF}]{{\small {\small
\begin{tikzpicture}
\begin{axis}[%
scale only axis,
width=0.23\textwidth,
height=0.15\textwidth,
ymin=0, ymax = 1,
xmin=20, 
xmax=90,
xtick={10,20,...,90},
xlabel={SI Reduction (dB)},
ylabel={},
xlabel near ticks,
ylabel near ticks,
axis on top,
legend entries={
	$\userDim = 12$ Outdoor , $\userDim = 12$ Indoor,
	$\userDim = 20$ Outdoor , $\userDim = 20$ Indoor,
	$\userDim = 28$ Outdoor , $\userDim = 28$ Indoor,,
},
legend style={at={(0.5,1.1)},anchor=south,nodes=right}
]
\addplot +[const plot, color=KeynoteBlue,  line width=.75pt, mark=none]
	plot [error bars/.cd, y dir=both, y explicit,  
		error bar style={line width=.75pt}, error mark options={rotate=90,line width=.5pt}]
	table {./DocData/SuppressionCDFs/SuppessionCDF-East-West-36Tx36Rx-Outdoor-12TxDims.dat};
\addplot +[color=KeynoteRed, line width=.75pt, mark=none]
	plot [error bars/.cd, y dir=both, y explicit,  
		error bar style={line width=.75pt}, error mark options={rotate=90,line width=.5pt}]
	table {./DocData/SuppressionCDFs/SuppessionCDF-East-West-36Tx36Rx-Indoor-12TxDims.dat};
\addplot +[const plot, color=KeynoteBlue, dashed,  line width=.75pt, mark=none]
	plot [error bars/.cd, y dir=both, y explicit,  
		error bar style={line width=.75pt}, error mark options={rotate=90,line width=.5pt}]
	table {./DocData/SuppressionCDFs/SuppessionCDF-East-West-36Tx36Rx-Outdoor-20TxDims.dat};
\addplot +[color=KeynoteRed, dashed, line width=.75pt, mark=none]
	plot [error bars/.cd, y dir=both, y explicit,  
		error bar style={line width=.75pt}, error mark options={rotate=90,line width=.5pt}]
	table {./DocData/SuppressionCDFs/SuppessionCDF-East-West-36Tx36Rx-Indoor-20TxDims.dat};
\addplot +[color=KeynoteBlue, dotted, line width=.75pt, mark=none]
	plot [error bars/.cd, y dir=both, y explicit,  
		error bar style={line width=.75pt}, error mark options={rotate=90,line width=.5pt}]
	table {./DocData/SuppressionCDFs/SuppessionCDF-East-West-36Tx36Rx-Outdoor-28TxDims.dat};
\addplot +[color=KeynoteRed, dotted, line width=.75pt, mark=none]
	plot [error bars/.cd, y dir=both, y explicit,  
		error bar style={line width=.75pt}, error mark options={rotate=90,line width=.5pt}]
	table {./DocData/SuppressionCDFs/SuppessionCDF-East-West-36Tx36Rx-Indoor-28TxDims.dat};
\end{axis}
\end{tikzpicture}
}}
	}
\caption{Self-interference reduction achieved by \softNull{} as a function of the number of \txDims\ preserved, $\userDim$.  Better self-interference reduction is observed outdoors, due to less backscattering, which leads to a more correlated self-interference channel matrix. \label{fig:suppression} }
\end{figure*}

%

\subsection{Impact of scattering on self-interference reduction}

\ashuOldNote{The phrase "preserved transmit dimensions" makes little sense at this point of the document. Perhaps because it is explained inadequately and there are multiple equivalent concepts in the paper.}\evanOldNote{Replace with ``effective antennas'' everywhere.}

The scattering environment has a significant impact on the performance of \softNull. 
We use the collected traces to study how the scattering environment impacts the tradeoff between self-interference reduction and \txDims\ achieved by \softNull. 
The 72-element array is used, with a $(\numTx, \numRx) = (36,36)$, East-West partition of the transmit and receive elements, as shown in Figure~\ref{fig:EastWestSplit}.  
Figure~\ref{fig:suppressionVsDofs} compares the tradeoff between self-interference reduction and preserved \txDims\ in the outdoor deployment versus the indoor deployment. 
The thin gray lines correspond to the self-interference reduction achieved for each of the $36$ antennas, while the thick red line corresponds to the self-interference reduction averaged over all $36$ antennas. 
Figure~\ref{fig:suppressionCDF} shows the empirical cumulative distribution function of the achieved self-interference reduction, both indoors and outdoors, for a selection of values for the number of \txDims\ preserved. 

We see in Figure~\ref{fig:suppressionVsDofs} that with all 36 \txDims\ preserved the self-interference is only suppressed (passively) by $20$~dB. 
But by giving up 16 \txDims\ and preserving $\userDim=20$ \txDims\ for the downlink, the self-interference is suppressed more than by $50$~dB. 
We also see in Figure~\ref{fig:suppressionVsDofs}, however, that the self-interference reduction in the indoor deployment is not nearly as good as in the outdoor deployment. To achieve $50$~dB self-interference reduction in the indoor deployment, 24 of the 36 \txDims{} must be given up (as opposed to 16 outdoors), leaving $\userDim=12$ for downlink transmission. The same array was used in both environments, the only difference being the backscattering environment. The reason for better performance outdoors than indoors is that the backscattering reduces the correlation of the self-interference among antennas that is present in a low scattering environment. Less correlation makes it harder to suppress the self-interference at multiple antennas without giving up more \txDims{}. 
More precisely, the \softNull{} precoder projects the transmit signal onto the $\userDim$ singular vectors corresponding to the smallest  $\userDim$ singular vectors. In other words, \softNull{} reduces self-interference by avoiding the $(\numTx -\userDim)$  dominant modes of the self-interference channel. Outdoors, the direct paths between antennas dominate any backscattered paths, leading to a more correlated self-interference matrix, and hence a large amount of the overall channel power resides in the dominant $(\numTx -\userDim)$ modes (singular values). Therefore very good self-interference reduction is acheived by avoiding just these first few dominant modes.  Indoors, however, multi-path backscattering tends to decorrelate the self-interference channel and thus leads to a more uniform distribution of power over the modes. Therefore, in the indoor environment less self-interference is suppressed by avoiding the  $(\numTx -\userDim)$ most dominant modes.

Figure~\ref{fig:suppressionCDF} shows the empirical cumulative distribution function of the achieved self-interference reduction, both indoors and outdoors, for a selection of values for the number of \txDims\ preserved, $\userDim$. We see in Figure~\ref{fig:suppressionCDF}  that for small $\userDim$ there is much more variation in the achieved self-interference reduction outdoors than there is indoors. For the outdoor deployment with $\userDim=12$, the self-interference reduction for a given antenna can be as much $90$~dB  and as little as $62$~dB, a $28$~dB difference. Indoors, however there is much less variation. For the indoor deployment with $\userDim=12$, the difference between best and worst self-interference reduction is only $10$~dB. 
More variation outdoors than indoors is also due to less backscattering outdoors than indoors.  
Outdoors, the backscattering is nearly nonexistent and direct paths between transmit and receive antenna dominate even for small $\userDim$. The characteristics of the direct-path self-interference channel seen by each receive antenna vary greatly.  For example, the 
receive antennas nearest the transmit antennas see less correlation among the transmit antennas (because of smaller angular spread) than the receive antennas farther away from the transmit antennas. Indoors, however, for smaller $\userDim$ the self-interference is dominated by backscattered paths. Unlike the direct paths, the characteristics of the direct-path self-interference channel seen by each receive antenna do not vary as much. Therefore for small $\userDim$, we expect to see more variation in self-interference reduction over the array outdoors that we see indoors.

\subsection{Achievable rate gains over half-duplex}
In the previous subsections, we observed that \softNull\ enables the array to reduce self-interference by giving up a fraction of the available \txDims\ for downlink, so that they may be used for self-interference reduction via beamforming. 
The question remains as to whether the gain in self-interference reduction is worth the cost of giving up the required \txDims{}. 
In particular, we wish to understand the scenarios in which \softNull\ can provide improved data rates over conventional half-duplex systems, and when \softNull\ cannot. 
The self-interference channels used in simulation are the self-interference channel traces collected as described in Section~\ref{sec:setup}. 
The uplink/downlink channel measurements are limited to 4 clients, therefore we use a mix of measured uplink/downlink channels and simulated uplink/downlink channels (especially in the case where more than four clients are considered). \clay{There seems to be a mix of future and present tense (and even past tense at the start of the paragraph).  Perhaps we should just remove all past/future and just say `we'?  (There are a lot of `will's scattered around.)} \evanNote{Good Point, done.}

Each trace consists of a $72\times72$ measurement of the self-interference channel,  a $72\times4$ measurements of the downlink channel to the mobile clients, and a  $4\times72$ measurement  of the uplink channel.  Each channel measurement is done over $64$ OFDM subcarriers in channel 4 (20~MHz band) of the 2.4~GHz ISM band.
We compute the uplink and downlink rates achievable for each trace by implementing the precoder and equalizer for the measured channels. We then compute achievable rates from the uplink and downlink signal-to-interference-plus-noise ratios achieved after equalization and precoding (assuming optimal channel codes).  
\revision{Since we are focusing in this study on small cells which use lower transmit power than macro cells, we assume a sum-power constraint at the array of $0$~dBm and a client constraint of $-10$~dBm. }
Also, we assume a receiver noise power of $-95$~dBm for both the array and client.

 \ashuNote{Let's use array and client} \evanNote{Done.}

When simulating half-duplex MU-MIMO, we allow the array to use all $\numTotal = 72$ transmit antennas, and constrain the array to divide time equally between uplink and downlink operation. The half-duplex MU-MIMO downlink uses the standard zero-forcing precoder and the uplink uses the standard linear decorrelator equalizer. 
When simulating \softNull\ we allow the array to operate the downlink and uplink simultaneously, but constrain the array to use $\numTx=36$ antennas that are for transmission only and $\numRx=36$ for reception. 
Based on the experiments and discussion in Section~\ref{sec:partition}, we choose the East-West partition of transmit and receive antennas shown in Figure~\ref{fig:EastWestSplit}. 
In simulation of \softNull, we set the downlink precoder, $\downPrecoder$ (applied before the \softNull), to be the standard zero-forcing precoder, 
after which we apply the \softNull{} precoder, $\selfPrecoder$. 
Receive-side digital cancellation of the self-interference is also simulated for \softNull. 
The effects of receiver limitations (in particularly AD quantization and LNA desensitization) are modeled via the dynamic range model of \cite{Day12FDMIMO, Day12FDRelay} (and references therein). In this dynamic range mode,  Gaussian noise is added to the received signal in proportion to the power of the received signal+interference \revision{(see Appendix~\ref{sec:dynamicRangeExample} for more explanation of this model)}. 
The constant of proportionality is the \emph{dynamic noise figure}, $D_0$, which we conservatively select to be $25$~dB. Setting the dynamic noise figure to $25$~dB effectively limits the amount of digital cancellation that can be achieved to no more that $25$~dB. Therefore all other self-interference reduction must come via the beamforming performed by the \softNull\ downlink precoder. 
We also compare against ``ideal full-duplex,'' where transmission and reception occurs at the same time but the self-interference is zero. 
Note that the ideal full-duplex rate will be less than twice the half-duplex rate, because we still assume that even for ideal full-duplex, $\numTx=36$ antennas that are for transmission  and $\numRx=36$ for reception, as opposed to half-duplex which uses all antennas for both transmission and reception, but in separate time slots. \revision{The details on how achievable rates are computed from the simulation results are given in Appendix~\ref{sec:computationDetails}}.

\subsubsection{Achievable rates for measured channels, $K$=4 clients}
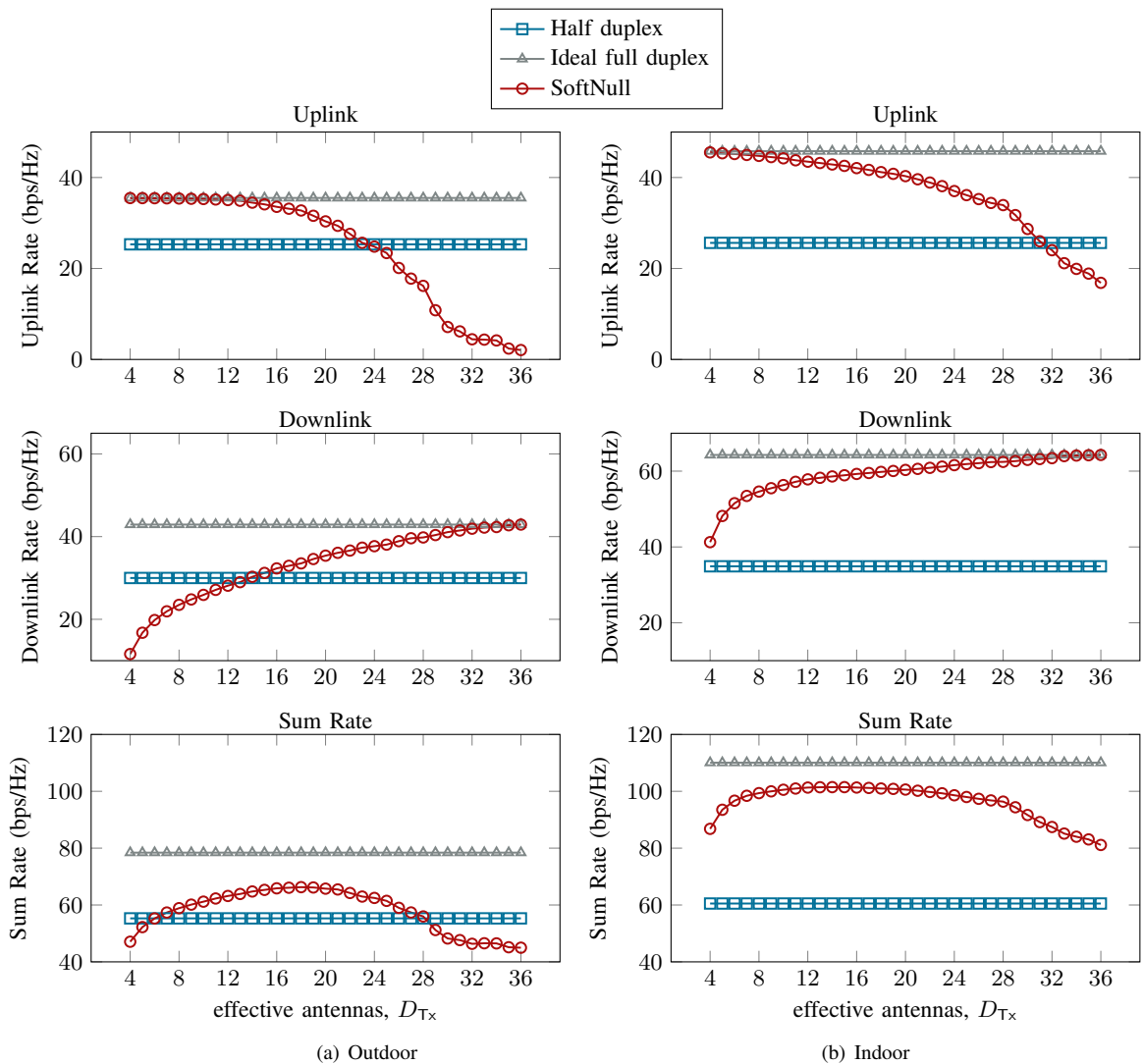
\begin{figure*}[htbp]
\centering
	\subfigure[Outdoor \label{fig:4KOutdoor}]{
	\centering
	{\small {\small
\begin{tikzpicture}
 \begin{groupplot}[
        group style={
            group name=my plots,
            group size=1 by 3,
            xlabels at=edge bottom,
            ylabels at=edge left,
        },
scale only axis,
width=0.35\textwidth,
height=0.17\textwidth,
xtick={4,8,...,36},
xlabel={\txDims, $\userDim$},
xlabel near ticks,
ylabel near ticks,
axis on top,
]
\nextgroupplot[ylabel={Uplink Rate (bps/Hz)},  title style={yshift=-7pt}, title={Uplink}, 
	legend style={at={(1.1,1.1)},anchor=south, nodes=right},
	legend entries={ Half duplex,  Ideal full duplex,  \softNull}, 
	ymin=0, ymax = 50
	]
\addplot [color=KeynoteBlue,  line width=.75pt, mark=square]
	plot [error bars/.cd, y dir=both, y explicit,  
		error bar style={line width=.75pt}, error mark options={rotate=90,line width=.5pt}]
	table [x = txDims, y = halfDuplexUplink,
		]{./\dataDir/Rates-East-West-36Tx36Rx-4K-Outdoor-InfdBPathLoss.dat};
\addplot [color=KeynoteGray,  line width=.75pt, mark=triangle]
	plot [error bars/.cd, y dir=both, y explicit,  
		error bar style={line width=.75pt}, error mark options={rotate=90,line width=.5pt}]
	table [x = txDims, y = idealDuplexUplink,
		]{./\dataDir/Rates-East-West-36Tx36Rx-4K-Outdoor-InfdBPathLoss.dat};
\addplot [color=KeynoteRed,  line width=.75pt, mark=o]
	plot [error bars/.cd, y dir=both, y explicit,  
		error bar style={line width=.75pt}, error mark options={rotate=90,line width=.5pt}]
	table [x = txDims, y = softNullUplink,
		]{./\dataDir/Rates-East-West-36Tx36Rx-4K-Outdoor-InfdBPathLoss.dat};
\nextgroupplot[ylabel={Downlink Rate (bps/Hz)}, title style={yshift=-7pt}, title={Downlink}, 
	ymin=10, ymax = 65
]
\addplot [color=KeynoteBlue,  line width=.75pt, mark=square]
	plot [error bars/.cd, y dir=both, y explicit,  
		error bar style={line width=.75pt}, error mark options={rotate=90,line width=.5pt}]
	table [x = txDims, y = halfDuplexDownlink,
		]{./\dataDir/Rates-East-West-36Tx36Rx-4K-Outdoor-InfdBPathLoss.dat};
\addplot [color=KeynoteGray,  line width=.75pt, mark=triangle]
	plot [error bars/.cd, y dir=both, y explicit,  
		error bar style={line width=.75pt}, error mark options={rotate=90,line width=.5pt}]
	table [x = txDims, y = idealDuplexDownlink,
		]{./\dataDir/Rates-East-West-36Tx36Rx-4K-Outdoor-InfdBPathLoss.dat};
\addplot [color=KeynoteRed,  line width=.75pt, mark=o]
	plot [error bars/.cd, y dir=both, y explicit,  
		error bar style={line width=.75pt}, error mark options={rotate=90,line width=.5pt}]
	table [x = txDims, y = softNullDownlink,
		]{./\dataDir/Rates-East-West-36Tx36Rx-4K-Outdoor-InfdBPathLoss.dat};
\nextgroupplot[ylabel={Sum Rate (bps/Hz)},title style={yshift=-7pt}, title={Sum Rate}, 
	ymin=40, ymax = 120
]
\addplot [color=KeynoteBlue,  line width=.75pt, mark=square]
	plot [error bars/.cd, y dir=both, y explicit,  
		error bar style={line width=.75pt}, error mark options={rotate=90,line width=.5pt}]
	table [x = txDims, y = halfDuplexSum,
		]{./\dataDir/Rates-East-West-36Tx36Rx-4K-Outdoor-InfdBPathLoss.dat};
\addplot [color=KeynoteGray,  line width=.75pt, mark=triangle]
	plot [error bars/.cd, y dir=both, y explicit,  
		error bar style={line width=.75pt}, error mark options={rotate=90,line width=.5pt}]
	table [x = txDims, y = idealDuplexSum,
		]{./\dataDir/Rates-East-West-36Tx36Rx-4K-Outdoor-InfdBPathLoss.dat};
\addplot [color=KeynoteRed,  line width=.75pt, mark=o]
	plot [error bars/.cd, y dir=both, y explicit,  
		error bar style={line width=.75pt}, error mark options={rotate=90,line width=.5pt}]
	table [x = txDims, y = softNullSum,
		]{./\dataDir/Rates-East-West-36Tx36Rx-4K-Outdoor-InfdBPathLoss.dat};
\end{groupplot}
\end{tikzpicture}
}}
	}\hspace{-2.5cm}
	\subfigure[Indoor \label{fig:4KIndoor}]{
	\centering
	{\small {\small
\begin{tikzpicture}
 \begin{groupplot}[
        group style={
            group name=my plots,
            group size=1 by 3,
            xlabels at=edge bottom,
            ylabels at=edge left,
        },
scale only axis,
width=0.35\textwidth,
height=0.17\textwidth,
xtick={4,8,...,36},
xlabel={\txDims, $\userDim$},
xlabel near ticks,
ylabel near ticks,
axis on top,
]
\nextgroupplot[ylabel={Uplink Rate (bps/Hz)},  title style={yshift=-7pt}, title={Uplink},
	ymin=0, ymax = 50
	]
\addplot [color=KeynoteBlue,  line width=.75pt, mark=square]
	plot [error bars/.cd, y dir=both, y explicit,  
		error bar style={line width=.75pt}, error mark options={rotate=90,line width=.5pt}]
	table [x = txDims, y = halfDuplexUplink,
		]{./\dataDir/Rates-East-West-36Tx36Rx-4K-Indoor-InfdBPathLoss.dat};
\addplot [color=KeynoteGray,  line width=.75pt, mark=triangle]
	plot [error bars/.cd, y dir=both, y explicit,  
		error bar style={line width=.75pt}, error mark options={rotate=90,line width=.5pt}]
	table [x = txDims, y = idealDuplexUplink,
		]{./\dataDir/Rates-East-West-36Tx36Rx-4K-Indoor-InfdBPathLoss.dat};
\addplot [color=KeynoteRed,  line width=.75pt, mark=o]
	plot [error bars/.cd, y dir=both, y explicit,  
		error bar style={line width=.75pt}, error mark options={rotate=90,line width=.5pt}]
	table [x = txDims, y = softNullUplink,
		]{./\dataDir/Rates-East-West-36Tx36Rx-4K-Indoor-InfdBPathLoss.dat};
\nextgroupplot[ylabel={Downlink Rate (bps/Hz)}, title style={yshift=-7pt}, title={Downlink},
		ymin=10, ymax = 70
]
\addplot [color=KeynoteBlue,  line width=.75pt, mark=square]
	plot [error bars/.cd, y dir=both, y explicit,  
		error bar style={line width=.75pt}, error mark options={rotate=90,line width=.5pt}]
	table [x = txDims, y = halfDuplexDownlink,
		]{./\dataDir/Rates-East-West-36Tx36Rx-4K-Indoor-InfdBPathLoss.dat};
\addplot [color=KeynoteGray,  line width=.75pt, mark=triangle]
	plot [error bars/.cd, y dir=both, y explicit,  
		error bar style={line width=.75pt}, error mark options={rotate=90,line width=.5pt}]
	table [x = txDims, y = idealDuplexDownlink,
		]{./\dataDir/Rates-East-West-36Tx36Rx-4K-Indoor-InfdBPathLoss.dat};
\addplot [color=KeynoteRed,  line width=.75pt, mark=o]
	plot [error bars/.cd, y dir=both, y explicit,  
		error bar style={line width=.75pt}, error mark options={rotate=90,line width=.5pt}]
	table [x = txDims, y = softNullDownlink,
		]{./\dataDir/Rates-East-West-36Tx36Rx-4K-Indoor-InfdBPathLoss.dat};
\nextgroupplot[ylabel={Sum Rate (bps/Hz)},title style={yshift=-7pt}, title={Sum Rate},
	ymin=40, ymax = 120
]
\addplot [color=KeynoteBlue,  line width=.75pt, mark=square]
	plot [error bars/.cd, y dir=both, y explicit,  
		error bar style={line width=.75pt}, error mark options={rotate=90,line width=.5pt}]
	table [x = txDims, y = halfDuplexSum,
		]{./\dataDir/Rates-East-West-36Tx36Rx-4K-Indoor-InfdBPathLoss.dat};
\addplot [color=KeynoteGray,  line width=.75pt, mark=triangle]
	plot [error bars/.cd, y dir=both, y explicit,  
		error bar style={line width=.75pt}, error mark options={rotate=90,line width=.5pt}]
	table [x = txDims, y = idealDuplexSum,
		]{./\dataDir/Rates-East-West-36Tx36Rx-4K-Indoor-InfdBPathLoss.dat};
\addplot [color=KeynoteRed,  line width=.75pt, mark=o]
	plot [error bars/.cd, y dir=both, y explicit,  
		error bar style={line width=.75pt}, error mark options={rotate=90,line width=.5pt}]
	table [x = txDims, y = softNullSum,
		]{./\dataDir/Rates-East-West-36Tx36Rx-4K-Indoor-InfdBPathLoss.dat};
\end{groupplot}
\end{tikzpicture}
}}
	}
\caption{Achievable rates of \softNull{} versus half-duplex as a function of the number of \txDims\ preserved, $\userDim$. An east-west $(\numTx, \numRx) = (36, 36)$ partition is considered. \softNull\ performs better indoors than outdoors because of less path loss.}
\label{fig:4KResult}
\end{figure*}

We assume the number of uplink and downlink clients are the same and both equal to four, $\numUplink=\numDownlink=K=4$. In this case we can directly use the uplink and downlink channel traces that were measured. 
Figure~\ref{fig:4KResult} shows the achievable uplink, downlink, and sum rates achieved by \softNull\ as a function of the chosen number of preserved \txDims, $\userDim$, and compares \softNull's performance to that of half-duplex as well as against ideal full-duplex.

First consider the results for the channels collected in the outdoor deployment, shown in Figure~\ref{fig:4KOutdoor}. 
The downlink rate achieved by \softNull\ increases as $\userDim$ increases, since more effective antennas become available to beamform and thus create a better signal-to-interference-plus-noise ratio to the downlink clients. 
However, as $\userDim$ increases the uplink rate decreases because \softNull\ can suppress less self-interference when more \txDims\ are used for downlink beamforming. 
Note that once $\userDim$ is less than approximately $12$ the incremental gain in uplink rate from giving up each additional effective antenna is negligible. For example,  at $\userDim=12$, the self-interference is sufficiently suppressed to no longer overwhelm the receiver, and digital cancellation removes remaining self-interference. 
Giving up more than 12 \txDims\ improves the uplink only slightly but greatly decreases the downlink rate.  
There is a range of values for $\userDim$ for which \softNull\ outperforms half-duplex both for the uplink and the downlink. 
The bottom plot of Figure~\ref{fig:4KOutdoor} shows the sum rate (uplink rate plus downlink rate). We see that \softNull\ outperforms half-duplex for $\userDim \in [5,28]$, achieving peak performance at $\userDim=18$. The achieved rate at $\userDim=18$ is $23\%$ better than half-duplex, but still $15\%$ less than the ideal full-duplex performance. 

Now consider the performance of \softNull{} for the indoor channel traces collected, shown in Figure~\ref{fig:4KIndoor}. 
In the indoor environment \softNull\ outperforms half-duplex for all values of $\userDim$, with the best performance coming at $\userDim=14$, for which a $62\%$ gain over half-duplex is achieved, but is still $12\%$ less than ideal full duplex. 

\begin{figure}[htbp]
\centering
	\subfigure[Outdoor\label{fig:partRatesOutdoor}]{
	{\small {\small
\begin{tikzpicture}
 \begin{groupplot}[
        group style={
            group name=my plots,
            group size=1 by 3,
            xlabels at=edge bottom,
            ylabels at=edge left,
        },
scale only axis,
width=0.30\textwidth,
height=0.20\textwidth,
xtick={4,8,...,36},
xlabel={\txDims, $\userDim$},
xlabel near ticks,
ylabel near ticks,
axis on top,
]
\nextgroupplot[ylabel={Sum Rate (bps/Hz)},
	legend style={at={(0.6,0.5)},anchor=south west,nodes=right}, 
	legend entries={East-West, North-South, NW-SE, Interleaved},
	ymin = 40, ymax = 120
	]
\addplot [color=KeynoteBlue,  line width=.75pt, mark=square]
	plot [error bars/.cd, y dir=both, y explicit,  
		error bar style={line width=.75pt}, error mark options={rotate=90,line width=.5pt}]
	table [x = txDims, y = softNullSum,
		]{./\dataDir/Rates-East-West-36Tx36Rx-4K-Outdoor-InfdBPathLoss.dat};
\addplot [color=KeynoteRed,  line width=.75pt, mark=diamond]
	plot [error bars/.cd, y dir=both, y explicit,  
		error bar style={line width=.75pt}, error mark options={rotate=90,line width=.5pt}]
	table [x = txDims, y = softNullSum,
		]{./\dataDir/Rates-North-South-36Tx36Rx-4K-Outdoor-InfdBPathLoss.dat};
\addplot [color=KeynoteYellow,  line width=.75pt, mark=o]
	plot [error bars/.cd, y dir=both, y explicit,  
		error bar style={line width=.75pt}, error mark options={rotate=90,line width=.5pt}]
	table [x = txDims, y = softNullSum,
		]{./\dataDir/Rates-Corners-36Tx36Rx-4K-Outdoor-InfdBPathLoss.dat};
\addplot [color=KeynoteGreen,  line width=.75pt, mark=x]
	plot [error bars/.cd, y dir=both, y explicit,  
		error bar style={line width=.75pt}, error mark options={rotate=90,line width=.5pt}]
	table [x = txDims, y = softNullSum,
		]{./\dataDir/Rates-Interleaved-36Tx36Rx-4K-Outdoor-InfdBPathLoss.dat};
\end{groupplot}
\end{tikzpicture}
}}
	}\hspace{-10pt}
	\subfigure[Indoor\label{fig:partRatesIndoor}]{
	{\small {\small
\begin{tikzpicture}
 \begin{groupplot}[
        group style={
            group name=my plots,
            group size=1 by 3,
            xlabels at=edge bottom,
            ylabels at=edge left,
        },
scale only axis,
width=0.30\textwidth,
height=0.20\textwidth,
xtick={4,8,...,36},
xlabel={\txDims, $\userDim$},
xlabel near ticks,
ylabel near ticks,
axis on top,
]
\nextgroupplot[ylabel={Sum Rate (bps/Hz)},
	ymin = 40, ymax = 120
		]
\addplot [color=KeynoteBlue,  line width=.75pt, mark=square]
	plot [error bars/.cd, y dir=both, y explicit,  
		error bar style={line width=.75pt}, error mark options={rotate=90,line width=.5pt}]
	table [x = txDims, y = softNullSum,
		]{./\dataDir/Rates-East-West-36Tx36Rx-4K-Indoor-InfdBPathLoss.dat};
\addplot [color=KeynoteRed,  line width=.75pt, mark=diamond]
	plot [error bars/.cd, y dir=both, y explicit,  
		error bar style={line width=.75pt}, error mark options={rotate=90,line width=.5pt}]
	table [x = txDims, y = softNullSum,
		]{./\dataDir/Rates-North-South-36Tx36Rx-4K-Indoor-InfdBPathLoss.dat};
\addplot [color=KeynoteYellow,  line width=.75pt, mark=o]
	plot [error bars/.cd, y dir=both, y explicit,  
		error bar style={line width=.75pt}, error mark options={rotate=90,line width=.5pt}]
	table [x = txDims, y = softNullSum,
		]{./\dataDir/Rates-Corners-36Tx36Rx-4K-Indoor-InfdBPathLoss.dat};
\addplot [color=KeynoteGreen,  line width=.75pt, mark=x]
	plot [error bars/.cd, y dir=both, y explicit,  
		error bar style={line width=.75pt}, error mark options={rotate=90,line width=.5pt}]
	table [x = txDims, y = softNullSum,
		]{./\dataDir/Rates-Interleaved-36Tx36Rx-4K-Indoor-InfdBPathLoss.dat};
\end{groupplot}
\end{tikzpicture}
}}
	}
\caption{\revision{Achievable rates for different antenna partitions. }}
\label{fig:partitionRates}
\end{figure}
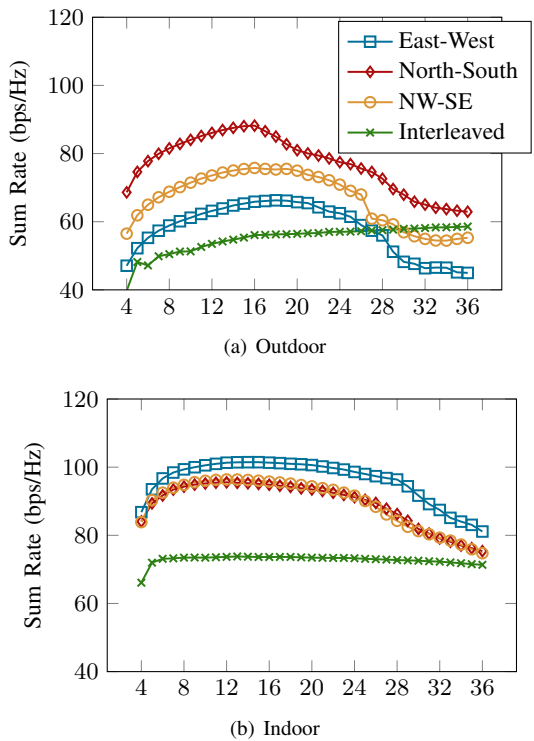

\revision{
Figure~\ref{fig:partitionRates} shows how the chosen antenna partition affects the achievable rate performance, in both the outdoor and indoor environment.  We know from Section~\ref{sec:partition} (Figure~\ref{fig:partitionsResults}) that contiguous partitions
(East-West, North-South, and Northwest-Southeast) all provide good suppression, with the East-West partition providing slightly more suppression than the others. The interleaved partition provides very poor suppression. Based on these results from Figure~\ref{fig:partitionsResults}, the achievable rate performance in the indoor environment, shown in Figure~\ref{fig:partRatesIndoor}, is expected: the East-West partition provides the best data rates, with the other contiguous partitions close behind, and the interleaved partition leading to poor data rates. 
The achievable rate performance in the outdoor environment is surprising, at first glance: we see in Figure~\ref{fig:partRatesOutdoor} that the North-South partition provides significantly better data rates than the other partitions, and the East-West partition (which has slightly better  suppression) gives the lowest data rates among the three contiguous partitions.  
In the outdoor experiment, the users were in line-of-sight of the array and at the same ground level as the array. In other words, the paths between the users and the array vary in azimuth but are approximately the same in elevation. Of the three contiguous partitions, the North-South partition has the longest horizontal dimension, and shortest vertical dimension, meaning that it  has the greatest azimuth resolution and the least elevation resolution. Because of the greater azimuth resolution, the vectors to each user are less aligned than they would be for other partitions, and thus the greatest SNR to each user can be achieved.  The East-West partition has the least azimuth resolution, and hence has the worst performance of the contiguous partitions for the outdoor test. In the indoor environment, because of multipath scattering, the paths between the array and the users have significant variation in both azimuth and elevation. Therefore in the indoor environment the array's resolution has much less impact on beamforming performance. Because the East-West partition has slightly better self-interference suppression, the East-West partition yields the highest data rates.  The take-away is that the choice of array partition depends not only on self-interference suppression performance, but also on matching the arrays'  angular resolution to the angular distribution of the users that will be served. 
}

At first, it seems surprising that the gains over half-duplex are \emph{better} indoors than outdoors, when we saw in Figure~\ref{fig:suppressionVsDofs} that the self-interference reduction achieved indoors is \emph{worse} than that achieved outdoors. 
The difference is that the path loss for the channels measured indoors was much less than that measured outdoors.  
The clients indoors were necessarily placed closer to the array \revisionTwo{(4-9 m)} because of limited space, but outdoors were placed much farther \revisionTwo{(9-15 m)}. 
Full-duplex always becomes more challenging as path loss increases. 

Larger path loss means the uplink signal is weaker, and therefore more self-interference reduction is required to make the self-interference commensurate in power to the uplink signal.  Hence, full-duplex systems to date have only been demonstrated for small cells \cite{Ashu12FDReview} \revision{(see Appendix~\ref{sec:dynamicRangeExample} for more explanation and a numerical example on how path loss and dynamic range limits the amount of required self-interference suppression)}. 
For \softNull\ in particular, larger path loss means more \txDims\ must be given up to achieve better self-interference reduction. Larger path loss also means that more \txDims\ are needed to achieve a sufficient signal strength on the downlink, therefore the cost of using \txDims\ for the sake of reducing self-interference is greater. Because the path loss was greater in the outdoor deployment than the indoor, the gains of \softNull{} are less for the outdoor deployment than for the indoor deployment.  Even though the achieved self-interference is better outdoors than indoors, the benefit of better suppression does not compensate for the greater path loss.

\revision{It is also interesting to note in Figure~\ref{fig:4KResult}, that although the ideal full-duplex rates are improved significantly indoors relative to outdoors, due to lower path loss, the half-duplex rates are only improved slightly indoors relative to outdoors. The reason is that half-duplex has a much larger beamforming gain than full-duplex, because it uses the full 72 antennas for transmission and reception, whereas ideal full-duplex only has 36 antennas (one partition for transmission, the other for reception). For half-duplex, with the large beamforming gain from 72 antennas, the received power is so large that the dominant noise source is not the thermal noise floor, which is independent of received power, but dynamic range limitations such as  quantization noise, oscillator noise, etc, which are \emph{proportional} to the received power (see Appendix~\ref{sec:dynamicRangeExample} for details of the dynamic range model). Lower path loss increases the received signal power, but also increases the noise due to dynamic range limitations by an equal amount, hence the modest improvement in SNR for lower path loss. For ideal full-duplex, the beamforming gain is much less since there are only 36 antennas, and received power is low enough for the thermal noise floor to be the dominant noise source. Lowering the path loss increases the received signal power relative the thermal noise floor, therefore significantly increasing the SNR, and leading to higher rates.}

We will next consider simulated uplink and downlink channels, so that path loss can be controlled for a fair comparison of outdoor versus indoor, and so that much larger path loss values can be considered.  

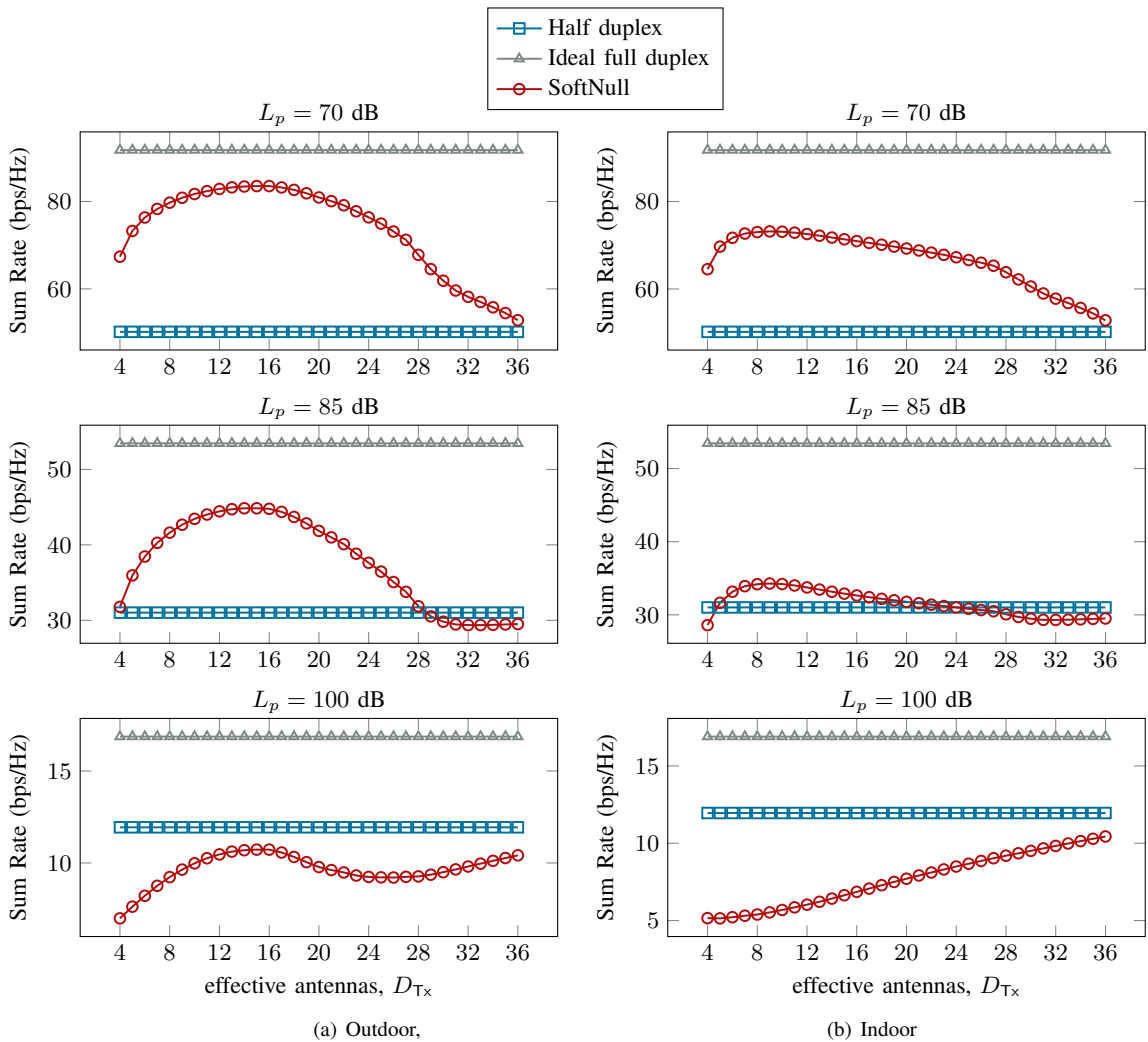
\begin{figure*}[htbp]
\centering
	\subfigure[Outdoor, \label{fig:4KOutdoor-varyPathLoss}]{
	\centering
	{\small {\small
\begin{tikzpicture}
 \begin{groupplot}[
        group style={
            group name=my plots,
            group size=1 by 3,
            xlabels at=edge bottom,
            ylabels at=edge left,
        },
scale only axis,
width=0.35\textwidth,
height=0.16\textwidth,
xtick={4,8,...,36},
xlabel={\txDims, $\userDim$},
xlabel near ticks,
ylabel near ticks,
axis on top,
]
\nextgroupplot[ylabel={Sum Rate (bps/Hz)},  title style={yshift=-7pt}, title={$L_p = 70$~dB}, 
	legend style={at={(1.1,1.1)},anchor=south, nodes=right},
	legend entries={ Half duplex,  Ideal full duplex,  \softNull}, 
	]
\addplot [color=KeynoteBlue,  line width=.75pt, mark=square]
	plot [error bars/.cd, y dir=both, y explicit,  
		error bar style={line width=.75pt}, error mark options={rotate=90,line width=.5pt}]
	table [x = txDims, y = halfDuplexSum,
		]{./\dataDir/Rates-East-West-36Tx36Rx-4K-Outdoor-70dBPathLoss.dat};
\addplot [color=KeynoteGray,  line width=.75pt, mark=triangle]
	plot [error bars/.cd, y dir=both, y explicit,  
		error bar style={line width=.75pt}, error mark options={rotate=90,line width=.5pt}]
	table [x = txDims, y = idealDuplexSum,
		]{./\dataDir/Rates-East-West-36Tx36Rx-4K-Outdoor-70dBPathLoss.dat};
\addplot [color=KeynoteRed,  line width=.75pt, mark=o]
	plot [error bars/.cd, y dir=both, y explicit,  
		error bar style={line width=.75pt}, error mark options={rotate=90,line width=.5pt}]
	table [x = txDims, y = softNullSum,
		]{./\dataDir/Rates-East-West-36Tx36Rx-4K-Outdoor-70dBPathLoss.dat};
\nextgroupplot[ylabel={Sum Rate (bps/Hz)},  title style={yshift=-7pt}, title={$L_p = 85$~dB}, 
	]
\addplot [color=KeynoteBlue,  line width=.75pt, mark=square]
	plot [error bars/.cd, y dir=both, y explicit,  
		error bar style={line width=.75pt}, error mark options={rotate=90,line width=.5pt}]
	table [x = txDims, y = halfDuplexSum,
		]{./\dataDir/Rates-East-West-36Tx36Rx-4K-Outdoor-85dBPathLoss.dat};
\addplot [color=KeynoteGray,  line width=.75pt, mark=triangle]
	plot [error bars/.cd, y dir=both, y explicit,  
		error bar style={line width=.75pt}, error mark options={rotate=90,line width=.5pt}]
	table [x = txDims, y = idealDuplexSum,
		]{./\dataDir/Rates-East-West-36Tx36Rx-4K-Outdoor-85dBPathLoss.dat};
\addplot [color=KeynoteRed,  line width=.75pt, mark=o]
	plot [error bars/.cd, y dir=both, y explicit,  
		error bar style={line width=.75pt}, error mark options={rotate=90,line width=.5pt}]
	table [x = txDims, y = softNullSum,
		]{./\dataDir/Rates-East-West-36Tx36Rx-4K-Outdoor-85dBPathLoss.dat};
\nextgroupplot[ylabel={Sum Rate (bps/Hz)},  title style={yshift=-7pt}, title={$L_p = 100$~dB},
	]
\addplot [color=KeynoteBlue,  line width=.75pt, mark=square]
	plot [error bars/.cd, y dir=both, y explicit,  
		error bar style={line width=.75pt}, error mark options={rotate=90,line width=.5pt}]
	table [x = txDims, y = halfDuplexSum,
		]{./\dataDir/Rates-East-West-36Tx36Rx-4K-Outdoor-100dBPathLoss.dat};
\addplot [color=KeynoteGray,  line width=.75pt, mark=triangle]
	plot [error bars/.cd, y dir=both, y explicit,  
		error bar style={line width=.75pt}, error mark options={rotate=90,line width=.5pt}]
	table [x = txDims, y = idealDuplexSum,
		]{./\dataDir/Rates-East-West-36Tx36Rx-4K-Outdoor-100dBPathLoss.dat};
\addplot [color=KeynoteRed,  line width=.75pt, mark=o]
	plot [error bars/.cd, y dir=both, y explicit,  
		error bar style={line width=.75pt}, error mark options={rotate=90,line width=.5pt}]
	table [x = txDims, y = softNullSum,
		]{./\dataDir/Rates-East-West-36Tx36Rx-4K-Outdoor-100dBPathLoss.dat};
\end{groupplot}
\end{tikzpicture}
}}
	} \hspace{-2.5cm}
	\subfigure[Indoor \label{fig:4KIndoor-varyPathLoss}]{
	\centering
	{\small {\small
\begin{tikzpicture}
 \begin{groupplot}[
        group style={
            group name=my plots,
            group size=1 by 3,
            xlabels at=edge bottom,
            ylabels at=edge left,
        },
scale only axis,
width=0.35\textwidth,
height=0.16\textwidth,
xtick={4,8,...,36},
xlabel={\txDims, $\userDim$},
xlabel near ticks,
ylabel near ticks,
axis on top,
]
\nextgroupplot[ylabel={Sum Rate (bps/Hz)},  title style={yshift=-7pt}, title={$L_p = 70$~dB}, 
	]
\addplot [color=KeynoteBlue,  line width=.75pt, mark=square]
	plot [error bars/.cd, y dir=both, y explicit,  
		error bar style={line width=.75pt}, error mark options={rotate=90,line width=.5pt}]
	table [x = txDims, y = halfDuplexSum,
		]{./\dataDir/Rates-East-West-36Tx36Rx-4K-Indoor-70dBPathLoss.dat};
\addplot [color=KeynoteGray,  line width=.75pt, mark=triangle]
	plot [error bars/.cd, y dir=both, y explicit,  
		error bar style={line width=.75pt}, error mark options={rotate=90,line width=.5pt}]
	table [x = txDims, y = idealDuplexSum,
		]{./\dataDir/Rates-East-West-36Tx36Rx-4K-Indoor-70dBPathLoss.dat};
\addplot [color=KeynoteRed,  line width=.75pt, mark=o]
	plot [error bars/.cd, y dir=both, y explicit,  
		error bar style={line width=.75pt}, error mark options={rotate=90,line width=.5pt}]
	table [x = txDims, y = softNullSum,
		]{./\dataDir/Rates-East-West-36Tx36Rx-4K-Indoor-70dBPathLoss.dat};
\nextgroupplot[ylabel={Sum Rate (bps/Hz)},  title style={yshift=-7pt}, title={$L_p = 85$~dB}, 
	]
\addplot [color=KeynoteBlue,  line width=.75pt, mark=square]
	plot [error bars/.cd, y dir=both, y explicit,  
		error bar style={line width=.75pt}, error mark options={rotate=90,line width=.5pt}]
	table [x = txDims, y = halfDuplexSum,
		]{./\dataDir/Rates-East-West-36Tx36Rx-4K-Indoor-85dBPathLoss.dat};
\addplot [color=KeynoteGray,  line width=.75pt, mark=triangle]
	plot [error bars/.cd, y dir=both, y explicit,  
		error bar style={line width=.75pt}, error mark options={rotate=90,line width=.5pt}]
	table [x = txDims, y = idealDuplexSum,
		]{./\dataDir/Rates-East-West-36Tx36Rx-4K-Indoor-85dBPathLoss.dat};
\addplot [color=KeynoteRed,  line width=.75pt, mark=o]
	plot [error bars/.cd, y dir=both, y explicit,  
		error bar style={line width=.75pt}, error mark options={rotate=90,line width=.5pt}]
	table [x = txDims, y = softNullSum,
		]{./\dataDir/Rates-East-West-36Tx36Rx-4K-Indoor-85dBPathLoss.dat};
\nextgroupplot[ylabel={Sum Rate (bps/Hz)},  title style={yshift=-7pt}, title={$L_p = 100$~dB}, 
	]
\addplot [color=KeynoteBlue,  line width=.75pt, mark=square]
	plot [error bars/.cd, y dir=both, y explicit,  
		error bar style={line width=.75pt}, error mark options={rotate=90,line width=.5pt}]
	table [x = txDims, y = halfDuplexSum,
		]{./\dataDir/Rates-East-West-36Tx36Rx-4K-Indoor-100dBPathLoss.dat};
\addplot [color=KeynoteGray,  line width=.75pt, mark=triangle]
	plot [error bars/.cd, y dir=both, y explicit,  
		error bar style={line width=.75pt}, error mark options={rotate=90,line width=.5pt}]
	table [x = txDims, y = idealDuplexSum,
		]{./\dataDir/Rates-East-West-36Tx36Rx-4K-Indoor-100dBPathLoss.dat};
\addplot [color=KeynoteRed,  line width=.75pt, mark=o]
	plot [error bars/.cd, y dir=both, y explicit,  
		error bar style={line width=.75pt}, error mark options={rotate=90,line width=.5pt}]
	table [x = txDims, y = softNullSum,
		]{./\dataDir/Rates-East-West-36Tx36Rx-4K-Indoor-100dBPathLoss.dat};
\end{groupplot}
\end{tikzpicture}
}}
	}
\caption{Achievable sum rates of \softNull{} versus half-duplex as a function of the number of \txDims\ preserved, $\userDim$, for several different values of path loss, $L_p$. An east-west $(\numTx, \numRx) = (36, 36)$ partition is considered. The larger the path loss, the more \txDims\ must be given up to achieve sufficient self-interference reduction. \softNull\ performance is better outdoors, because of better self-interference reduction due to more correlation in the self-interference channel. \label{fig:4KResult-varyPathLoss}}
\end{figure*}

Figure~\ref{fig:4KResult-varyPathLoss} compares the outdoor and indoor performance of \softNull\ for  path loss values of $70$, $85$ and $100$~dB.  
At 2~GHz these values correspond roughly to distances of 50, 300, and 1000~m in outdoor line-of-sight conditions (i.e. assuming path loss exponent of 2). Indoors, these path loss values correspond to distances of 3, 10, and 30~m for indoor non line-of-sight conditions (i.e. assuming a path loss exponent of 3) \cite{Rappaport}.	 
The self-interference channels are the channels measured indoors and outdoors, but the uplink and downlink channels are i.i.d. Rayleigh distributed channels, with a controlled path loss. 
Both outdoor and indoor, as the path loss increases the gain of \softNull\ over half-duplex decreases. 
More path loss means that more self-interference must be suppressed to make self-interference commensurate to the uplink received signal strength, therefore we see in Figure~\ref{fig:4KResult-varyPathLoss} that as the path loss increases, the optimal value of $\userDim$ decreases in order to provide the needed self-interference reduction. 
Also, as expected, when the path loss is the same in both environments, the performance is better outdoors than indoors, because (as discussed above) \softNull\ can better reduce self-interference outdoors than indoors. 
For the outdoor deployment, \softNull\ can significantly outperform half-duplex at 70 and 85~dB paths loss,  but at at 100~dB path loss, \softNull\ cannot outperform half-duplex for any value of $\userDim$. At 100~dB path loss indoors, too many \txDims\ must be given up to achieve the required self-interference reduction, and the downlink rate suffers accordingly. Similarly in the indoor environment, where full-duplex is more challenging because of less self-interference suppression achieved, \softNull\ can significantly outperform half-duplex at 70~dB path loss, just barely outperforms half-duplex at 85~dB paths loss, and underperforms half-duplex at 100~dB path loss. 

\subsection{Varying number of clients}
\begin{figure*}[hbtp]
\centering
	\subfigure[Outdoor]{
	\centering
	{\small {\small
\begin{tikzpicture}
\begin{axis}[%
scale only axis,
width=0.30\textwidth,
height=0.20\textwidth,
xtick={4,8,...,36},
xlabel={Number of users, $K$},
ylabel={Achievable Rate (bps/Hz)},
xlabel near ticks,
ylabel near ticks,
axis on top,
legend entries={
	Half Duplex $L_p=70$~dB, \softNull\ $L_p=70$~dB, 
	Half Duplex $L_p=85$~dB, \softNull\ $L_p=85$~dB,
	Half Duplex $L_p=100$~dB\ \ , \softNull\ $L_p=100$~dB,
},
legend columns = 2, 
legend style={at={(1.1,1.1)},anchor=south, nodes=right}
]
\addplot [color=KeynoteBlue,  line width=.75pt, mark=square*, mark options={solid}, dashed]
	plot [error bars/.cd, y dir=both, y explicit,  
		error bar style={line width=.75pt}, error mark options={rotate=90,line width=.5pt}]
	table [x = txDims, y = HalfDuplex,
		]{./\dataDir/VaryNumUsers-East-West-36Tx36Rx-Outdoor-70dBPathLoss.dat};
\addplot [color=KeynoteRed,  line width=.75pt, mark=square*]
	plot [error bars/.cd, y dir=both, y explicit,  
		error bar style={line width=.75pt}, error mark options={rotate=90,line width=.5pt}]
	table [x = txDims, y = SoftNull,
		]{./\dataDir/VaryNumUsers-East-West-36Tx36Rx-Outdoor-70dBPathLoss.dat};
\addplot [color=KeynoteBlue,  line width=.75pt, mark=diamond*, mark options={solid}, dashed]
	plot [error bars/.cd, y dir=both, y explicit,  
		error bar style={line width=.75pt}, error mark options={rotate=90,line width=.5pt}]
	table [x = txDims, y = HalfDuplex,
		]{./\dataDir/VaryNumUsers-East-West-36Tx36Rx-Outdoor-85dBPathLoss.dat};
\addplot [color=KeynoteRed,  line width=.75pt, mark=diamond*]
	plot [error bars/.cd, y dir=both, y explicit,  
		error bar style={line width=.75pt}, error mark options={rotate=90,line width=.5pt}]
	table [x = txDims, y = SoftNull,
		]{./\dataDir/VaryNumUsers-East-West-36Tx36Rx-Outdoor-85dBPathLoss.dat};
\addplot [color=KeynoteBlue,  line width=.75pt, mark=* , mark options={solid}, dashed]
	plot [error bars/.cd, y dir=both, y explicit,  
		error bar style={line width=.75pt}, error mark options={rotate=90,line width=.5pt}]
	table [x = txDims, y = HalfDuplex,
		]{./\dataDir/VaryNumUsers-East-West-36Tx36Rx-Outdoor-100dBPathLoss.dat};
\addplot [color=KeynoteRed,  line width=.75pt, mark=*]
	plot [error bars/.cd, y dir=both, y explicit,  
		error bar style={line width=.75pt}, error mark options={rotate=90,line width=.5pt}]
	table [x = txDims, y = SoftNull,
		]{./\dataDir/VaryNumUsers-East-West-36Tx36Rx-Outdoor-100dBPathLoss.dat};
\end{axis}
\end{tikzpicture}
}}
	}\hspace{-4.5cm}
	\subfigure[Indoor]{
	\centering
	{\small {\small
\begin{tikzpicture}
\begin{axis}[%
scale only axis,
width=0.30\textwidth,
height=0.20\textwidth,
xtick={4,8,...,36},
xlabel={Number of users, $K$},
ylabel={Achievable Rate (bps/Hz)},
xlabel near ticks,
ylabel near ticks,
axis on top,
]
\addplot [color=KeynoteBlue,  line width=.75pt, mark=square*, mark options={solid}, dashed]
	plot [error bars/.cd, y dir=both, y explicit,  
		error bar style={line width=.75pt}, error mark options={rotate=90,line width=.5pt}]
	table [x = txDims, y = HalfDuplex,
		]{./\dataDir/VaryNumUsers-East-West-36Tx36Rx-Indoor-70dBPathLoss.dat};
\addplot [color=KeynoteRed,  line width=.75pt, mark=square*]
	plot [error bars/.cd, y dir=both, y explicit,  
		error bar style={line width=.75pt}, error mark options={rotate=90,line width=.5pt}]
	table [x = txDims, y = SoftNull,
		]{./\dataDir/VaryNumUsers-East-West-36Tx36Rx-Indoor-70dBPathLoss.dat};
\addplot [color=KeynoteBlue,  line width=.75pt, mark=diamond*, mark options={solid}, dashed]
	plot [error bars/.cd, y dir=both, y explicit,  
		error bar style={line width=.75pt}, error mark options={rotate=90,line width=.5pt}]
	table [x = txDims, y = HalfDuplex,
		]{./\dataDir/VaryNumUsers-East-West-36Tx36Rx-Indoor-85dBPathLoss.dat};
\addplot [color=KeynoteRed,  line width=.75pt, mark=diamond*]
	plot [error bars/.cd, y dir=both, y explicit,  
		error bar style={line width=.75pt}, error mark options={rotate=90,line width=.5pt}]
	table [x = txDims, y = SoftNull,
		]{./\dataDir/VaryNumUsers-East-West-36Tx36Rx-Indoor-85dBPathLoss.dat};
\addplot [color=KeynoteBlue,  line width=.75pt, mark=* , mark options={solid}, dashed]
	plot [error bars/.cd, y dir=both, y explicit,  
		error bar style={line width=.75pt}, error mark options={rotate=90,line width=.5pt}]
	table [x = txDims, y = HalfDuplex,
		]{./\dataDir/VaryNumUsers-East-West-36Tx36Rx-Indoor-100dBPathLoss.dat};
\addplot [color=KeynoteRed,  line width=.75pt, mark=*]
	plot [error bars/.cd, y dir=both, y explicit,  
		error bar style={line width=.75pt}, error mark options={rotate=90,line width=.5pt}]
	table [x = txDims, y = SoftNull,
		]{./\dataDir/VaryNumUsers-East-West-36Tx36Rx-Indoor-100dBPathLoss.dat};
\end{axis}
\end{tikzpicture}
}}
	}
\caption{Achievable rate of SoftNull and half-duplex as function of the number of clients being served, $K$. Results are for array with $
M=72$, and $(\numTx, \numRx) = (36,36)$, both outdoor (a) and indoor (b).  For low number of clients \softNull\ outperforms half-duplex, and for larger number of clients  \softNull\ underperforms half-duplex operation.  \ashuNote{Plot \% gain too?} \evanNote{We could, would be redundant in my opinion, however.}}
\label{fig:numUsersResult}
\end{figure*}
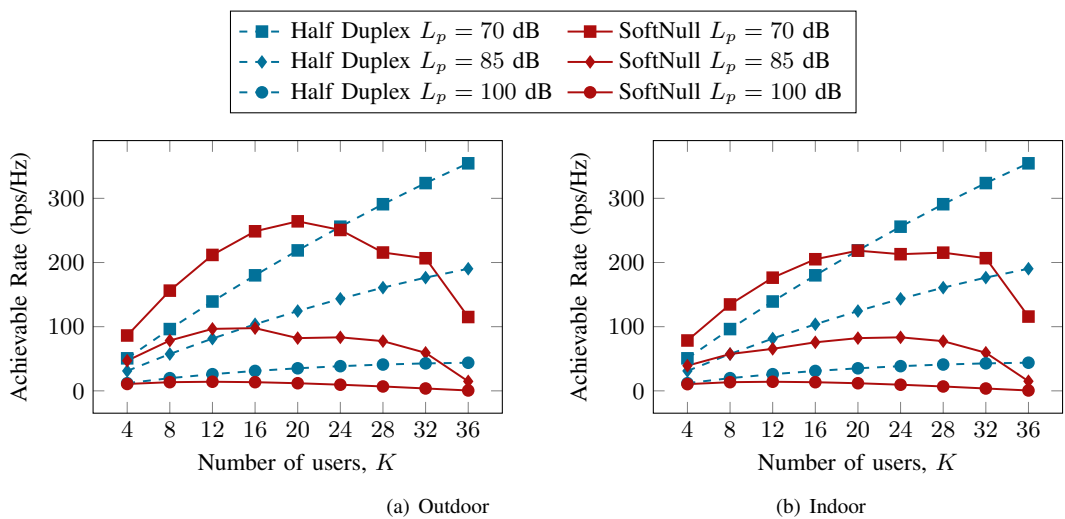

Figure~\ref{fig:numUsersResult} shows  how the performance of \softNull\ compares to that of half-duplex as a function of the number of uplink clients and downlink clients (which are assumed to be equal). 
For half-duplex, adding clients is a opportunity to provide higher sum rates via spatial multiplexing. 
Likewise for full-duplex with \softNull{}, adding more clients provides a multiplexing opportunity, but adding more clients also means that fewer \txDims\ can be given up for the sake of self-interference reduction. 
For each of the data points plotted for \softNull, we assume \softNull\ selects the $\userDim$ value that maximizes the sum rate. 
The simulations were carried out for path loss values of 70, 85, and 100~dB. 
We see in Figure~\ref{fig:numUsersResult} that \softNull\ outperforms half-duplex for small numbers of clients (except when the path loss is large). However, as more clients are added, eventually a point is reached at which \softNull\ underperforms half-duplex.   
\revisionTwo{Half-duplex benefits from the fact all 72 antennas can be used for uplink or downlink transmission and that no effective antennas must be sacrificed for self-interference suppression. Therefore, in systems where the number of users served is similar to the number of antennas (i.e. when $K \approx M$), the spatial-multiplexing benefit from having access to more antennas in half-duplex outweighs the time-multiplexing benefit from transmitting and receiving simultaneously in full-duplex via \softNull.  However, in systems where the number of users served in a given time slot is small relative to the number of antennas (i.e, when $K \ll M$) -- the regime-of-interest for many-antenna wireless (also called ``massive MIMO'') \cite{Marzetta2010NoCooperativeMassiveMIMO, Marzetta14MassiveMIMOOverview, Rusek12MassiveMimoOverview} -- the benefit of \softNull\ is more pronounced. 
We see in~Figure~\ref{fig:numUsersResult} that the threshold on $K$ below which \softNull\   outperforms full-duplex depends heavily on the path loss. }
Consider the outdoor case, shown in Figure~\ref{fig:numUsersResult}. 
When the path loss is only $70$~dB, serving more clients increases \softNull\ performance up to $K\leq20$, after which \softNull\ cannot serve more clients without allowing prohibitive self-interference.  Therefore the rate decreases for $K>20$, and \softNull\ underperforms half-duplex for $K>20$ clients. 
As the path loss increases, more \txDims\ must be given up to sufficiently suppress self-interference and avoid swamping the uplink signal, and hence fewer clients can be served. 
For $85$~dB path loss, \softNull\ outperforms half-duplex when $K\leq12$. 
And in the $100$~dB path loss case \softNull\ underperforms half-duplex for each value of $K$ simulated. 
In the indoor environment, more effective antennas must be used to achieve the same self-interference reduction, therefore the value for $K$ at which \softNull\ outperforms half-duplex is smaller. 
For $70$~dB path loss \softNull\ outperforms half-duplex for $K\leq16$, and for $85$~dB path loss \softNull\ outperforms half-duplex for $K\leq4$. 
At 100~dB path loss, \softNull\ strictly underperforms half-duplex in the indoor deployment.  

\section{Conclusion}
\label{sec:conclusion}
\softNull\ provides an opportunity to enable full-duplex operation with current base station radios without requiring additional circuitry for analog cancellation. 
The primary intuition behind the \softNull\ precoder is that the self-interference need not be perfectly nulled; we only need to sacrifice the minimum number of \txDims\ required to sufficiently suppress self-interference. 
Our analysis based on the channels measured using a 72-element array shows that when the path loss is not too large, sufficient self-interference reduction can be achieved while only using a portion of the \txDims\ for self-interference suppression. 
However, due to additional backscattering indoors, more \txDims\ must be used for self-interference reduction indoors to achieve the same level of self-interference reduction outdoors. 
\softNull\ also performs quite well in the regime where $\numTx \gg K$. We note that the trend in wireless deployments is towards larger arrays, and  \softNull\ can provide a new full-duplex transmission mode using only half-duplex hardware. We would like to point the readers to a extension of this work in the related thesis \cite{Everett15PhDThesis}, in which a software-defined radio prototype of \softNull{} is presented. 

\appendices

\section{Optimal precoder solution}
\label{sec:precoderProof}

The optimal precoder is the solution to the  optimization problem
\begin{align}
	\selfPrecoder =& \underset{\precoder}{\argmin} \|\selfChannel \precoder \|_F^2,  
	\\ &\text{subject to } {\herm{\precoder}\precoder = \identity{\userDim}{\userDim}}.
\label{eq:problem1}
\end{align}
Via simple manipulations, the optimization function in \label{eq:problem1} can be rewritten as 
\begin{align}
	\label{eq:problem2}
	\selfPrecoder
	=&\ \underset{\precoder}{\argmin}\  \frac{1}{2} \trace\left(\herm{\precoder}\herm{\selfChannel} \selfChannel\precoder\right), \\ & \text{subject to } {\herm{\precoder}\precoder = \identity{\userDim}{\userDim}}.
\end{align}
Problem~(\ref{eq:problem2}) is a convex optimization problem with equality constraints, and can thus be solved by the method of Lagrange multipliers. The Lagrangian for Problem~(\ref{eq:problem2}) is
\begin{align}
L\left( \precoder, \lagMult \right) &= \frac{1}{2} \trace\left(\herm{\precoder}\herm{\selfChannel} \selfChannel\precoder\right) -
\frac{1}{2} \trace \left( \lagMult \herm{\precoder}\precoder \right), \nonumber
\\ & \herm{\lagMult} = \lagMult
\label{eq:lagrange}
\end{align}
The gradient of the Lagrangian with respect to $\precoder$ is 
$
\nabla_{\precoder} L\left( \precoder, \lagMult \right) = \herm{\selfChannel} \selfChannel \precoder - \precoder \lagMult.
$
Therefore the stationary points of the Lagrangian must satisfy
\begin{align}
 \herm{\selfChannel} \selfChannel \precoder &= \precoder \lagMult 
 \label{eq:eigenForm}
\end{align}
Let $\{\eigval_i\}_{i=1}^{\numTx}$ and $\{\eigvec_i\}_{i=1}^{\numTx}$ denote the eigenvalues and eigenvectors of $\herm{\selfChannel} \selfChannel$, and we assume that the set of eigenvalues is ordered from largest to smallest $\eigval_i$.
It is easy to check that (\ref{eq:eigenForm}) is satisfied if  $\lagMult$ is diagonal, and  the $j$th column of $\precoder$ is an eigenvector of $ \herm{\selfChannel} \selfChannel $, with the $j$th diagonal element of $\lagMult$ set to the corresponding eigenvalue. Note that the columns of  $\precoder$ need not be distinct to satisfy (\ref{eq:eigenForm}). For example, the columns of  $\precoder$ could each be the same eigenvector. However, $\precoder$ must also satisfy the constraint ${\herm{\precoder}\precoder = \identity{\userDim}{\userDim}}$. Note that since $ \herm{\selfChannel} \selfChannel $ is a hermitian $\numTx \times \numTx$ matrix, there exists a set of $\numTx$ orthonormal eigenvectors of $\herm{\selfChannel} \selfChannel$. Hence, ${\herm{\precoder}\precoder = \identity{\userDim}{\userDim}}$ is satisfied when the $\userDim$ columns of $\precoder$ consist of $\userDim$ \emph{distinct}  eigenvectors of $ \herm{\selfChannel} \selfChannel$. Therefore, any matrix $\precoder$ whose columns are distinct eigenvectors of $ \herm{\selfChannel} \selfChannel$ is a both a feasible point and a stationary point of the Lagrangian and thus a candidate for an optimal solution.
It remains to determine which choice of eigenvectors for the columns of $\precoder$ leads to the smallest value of the objective function $\frac{1}{2} \trace\left(\herm{\precoder}\herm{\selfChannel} \selfChannel\precoder\right)$.  It is fairly obvious that the eigenvectors corresponding to the smallest eigenvalues will lead to the optimal value, but we include the details below. 
Let $\mathcal{P} \subset \{1,2,\dots,\numTx\}$ denote the set of $\userDim$ indices corresponding to which eigenvectors of $\herm{\selfChannel} \selfChannel$ constitute the columns of $\precoder$.
One can check that  
\begin{equation}
\frac{1}{2} \trace\left(\herm{\precoder}\herm{\selfChannel} \selfChannel\precoder\right) = \frac{1}{2}\sum_{i\in \mathcal{P}}\lambda_i,
\end{equation}
which is minimized when $\mathcal{P}= \{\numTx-\userDim+1, \numTx-\userDim+2, \dots, \numTx\}$, the indices corresponding to the smallest eigenvalues of $\herm{\selfChannel} \selfChannel$. Therefore, the solution to Problem (\ref{eq:problem2}) is
\begin{equation}
\label{eq:svdPrecoderProved}
\selfPrecoder = \left[ \singVecRight{\numTx-\userDim+1}, \singVecRight{\numTx-\userDim+2}, \dots, \singVecRight{\numTx}\right].
\end{equation} Note that the eigenvectors of $\herm{\selfChannel} \selfChannel$ are equal to the right singular vectors of $\selfChannel$,  where  $\selfChannel = \singLeft \singVals \herm{\singRight}$, is the singular value decomposition of $\selfChannel$.
Therefore, an equivalent characterization of the optimal solution is that the columns of $\selfChannel$ are drawn from the right singular vectors of $\selfChannel$ corresponding to the $\userDim$ smallest singular values.

%


\revision{
\section{Rate Computation Details}
\label{sec:computationDetails}
Uplink and downlink achievable rates are computed as follows. We use the common method of computing ergodic achievable rates from the per-packet signal-to-interference-plus-noise ratios  \cite{TseBook, Duarte10Beamforming, Naren2010Beamforming}. 
Let the variable $\scheme \in \{\HD, \SN, \IFD \}$  be an index into the set of schemes we are comparing: half-duplex, \softNull{}, and ideal full-duplex. Let $\downFraction{\scheme} \in [0,1]$ 
denote the fraction of time scheme $\scheme$ allocates for downlink communication, and likewise $\upFraction{\scheme} \in [0,1]$  denote the fraction of time allocated for uplink. Obviously, since the schemes ideal-duplex and \softNull{} are full-duplex schemes in which uplink and downlink are always simultaneously active, 
$\downFraction{\IFD} = \upFraction{\IFD} = 1$
and 
$\downFraction{\SN} = \upFraction{\SN} = 1.$  However, for half-duplex, uplink and downlink must occur on separate time slots (or frequency bands). We assume an even allocation between uplink time and downlink time, that is
$\downFraction{\HD} = \upFraction{\HD} = \frac{1}{2}. $
When simulating scheme $\scheme$, for each channel realization $p$,  
we measure the  signal-to-interference-plus-noise ratio at each downlink user $j$, $\DownSINR{\scheme}(p,j)$. The average ergodic achievable downlink rate is computed by summing over the downlink users and averaging over channel realizations
\begin{equation}
\DownRate{\scheme} =  \downFraction{\scheme} \frac{1}{N_p} \sum_{p = 1}^{N_p} \sum_{j = 1}^{\numDownlink} \log_2[1+\DownSINR{\scheme}(p,j)], \label{eq:DwnRate}
\end{equation}
where $N_p$ is the number of channel realizations.
Likewise, the average ergodic achievable uplink rate is computed by summing the rates over the uplink users and averaging over channel realizations
\begin{equation}
\UpRate{\scheme}  = \upFraction{\scheme} \frac{1}{N_p} \sum_{p = 1}^{N_p} \sum_{i = 1}^{\numUplink} \log_2[1+\UpSINR{\scheme}(p,j)]. \label{eq:UpRate}
\end{equation}
The uplink SINR for \softNull, $\UpSINR{\SN}$ will be much smaller than $\UpSINR{\HD}$ because of residual self-interference.  Similarly $\DownSINR{\SN}$ will be smaller than $\DownSINR{\HD}$ because of effective transmit antennas sacrificed by \softNull{} for self-interference suppression. However, \softNull{} can still outperform half-duplex since \softNull{} due to the multiplexing gain of operating uplink and downlink concurrently, captured by setting $\downFraction{\SN} = \upFraction{\SN} = 1$ whereas $\downFraction{\HD} = \upFraction{\HD} = \frac{1}{2}$.

We note that Equations~\ref{eq:DwnRate}~and~\ref{eq:UpRate} are based on two assumptions: (1) that optimal channel codes are used, such that the ideal Shannon rate is achieved, and (2)  that the channel codes can extend over multiple channel realizations. Although these assumptions may not hold in practice, they provide a fair way of comparing system performance among different schemes, without the results being particular to a certain coding scheme and modulation rate. The sum rates are simply the sum of the uplink and downlink rates explained above. 
}

\revision{
\section{Explanation and Example of Dynamic Range Limitation}
\label{sec:dynamicRangeExample}
In wireless communication theory, one often assumes a fixed noise floor, based on thermal noise. 
For example, one could assume a thermal noise floor of $-90$~dBm, and since the noise floor is fixed, if the received signal power is $-80$~dBm, we would say the SNR is $10$~dB, and if the received signal power is $-70$~dBm we would say the SNR is $20$~dB.  However, the model of a fixed noise floor, is only true for low SNR. In practice there are noise sources whose power \emph{scales proportionally to the received power level}, such as A/D quantization noise, oscillator noise, amplifier non-linearities, etc \cite{Day12FDRelay}. Collectively, such impairments are called \emph{dynamic range limitation}. One way to effectively model dynamic range limitation is by adding a gaussian noise term whose power level is proportional to the received signal power \cite{Day12FDMIMO,Gray93DynamicRange, Namgoog05DynamicRange}, i.e., having a ``dynamic range noise floor'' that is fixed \emph{relative} to the received power. For example, a common assumption is that the dynamic range floor is $30$~dB below the received power level. Consider our example of a  $-90$~dBm thermal noise floor, and a dynamic range noise floor which is $40$~dB below the received power level. This means that for received power levels less than $-50$~dBm, dynamic range limitations have little effect, because thermal noise is the dominant noise source. But if the received signal power level is more than $-50$~dBm, the dynamic range floor becomes the dominant noise source, limiting the SNR to be no more than $40$~dB, even for very high received signal power.

Now let us consider the impact of limited dynamic range on full-duplex in a small cell, for the same example given above where the dynamic range noise floor which is $40$~dB below the received power level. Let us assume the base station transmits at $0$~dBm and the uplink users also transmit with power $0$~dBm, and let us further assume that there is $80$~dB path loss between the base station and the user(s). In this example the power of the desired uplink signal received at the base station will be $-80$~dBm. Now let's say that the self-interference reduction provided by the base station is only $20$~dB (this is about what could be achieved by merely separating the receive and transmit antennas by a foot).  Then the self-interference received by the base station is at $-20$~dBm. The self-interference dominates the received power level, and raises the dynamic range noise floor to $-20\dBm - 40\dB = -60\dBm.$ Even if the residual self-interference is digitally cancelled perfectly, the SNR for the desired uplink signal is $$-80\dBm - (-60\dBm) = -20\dB,$$ which much too low of an SNR to support a useful communication link. However if the amount of self-interference reduction is improved from $20$~dB to $50$~dB, then the received power level is reduced to $-50\dBm,$ and thus the dynamic range noise floor is reduced to $-50\dBm - 40\dB  = -90\dBm.$ Now, after the the residual self-interference is cancelled digitally, the the SNR of the uplink signal is $$-80\dBm - (-90\dBm) = 10\dB,$$ a positive SNR, which means that a useful communication link can be operated on the uplink in conjunction with downlink transmission.

We remark that the amount of self-interference suppression required achieve a positive uplink SNR is  dependent on the \emph{path loss} between the base station and the uplink user(s). If the path loss had been $100\dB$ instead of $80\dB$, then $70\dB$ of suppression, rather than $50\dB$, would be required to achieve an uplink SNR of $10\dB$. That is, the amount of required self-interference suppression is proportional to the amount of path loss between base station and users.  
}

\bibliographystyle{IEEEtran}
\bibliography{IEEEabrv,DocBib/EvanBib,DocBib/ClayBiB} 

\begin{IEEEbiography}[{\includegraphics[width=1in,height=1.25in,clip,keepaspectratio]{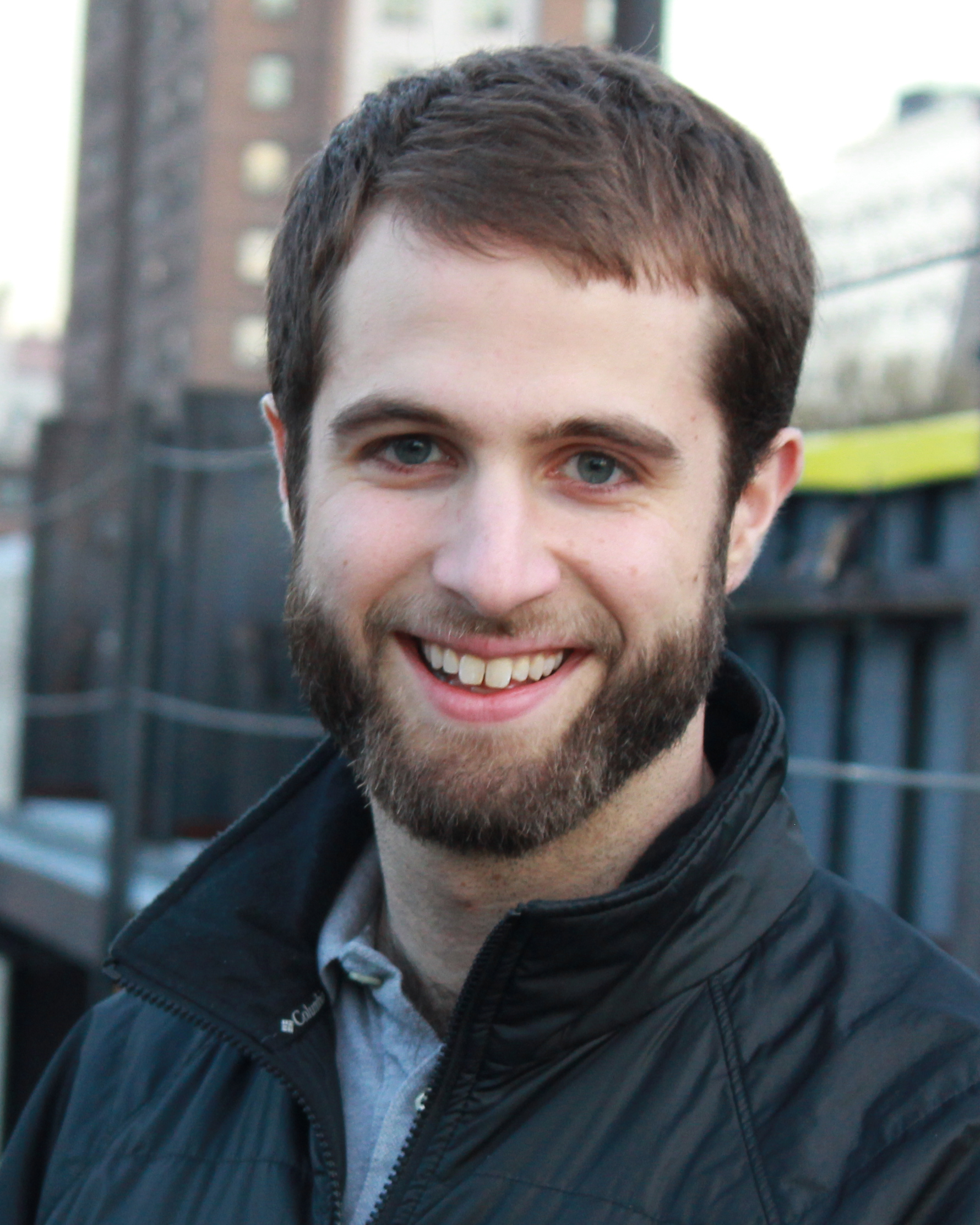}}]{Evan Everett} received the B.Eng degree in Wireless Engineering and the B.S degree in Physics from Auburn University in 2010. He received the Ph.D degree in Electrical Engineering from Rice University in 2016. He is currently a research scientist with Numerica Corp. in Fort Collins, CO. His research focus is theory and design for full-duplex wireless communication. He received the 2010 Comer Award for Excellence in Physical Science from Auburn University, and he is a National Science Foundation Graduate Research Fellow.\end{IEEEbiography}

\begin{IEEEbiography}[{\includegraphics[width=1in,height=1.25in,clip,keepaspectratio]{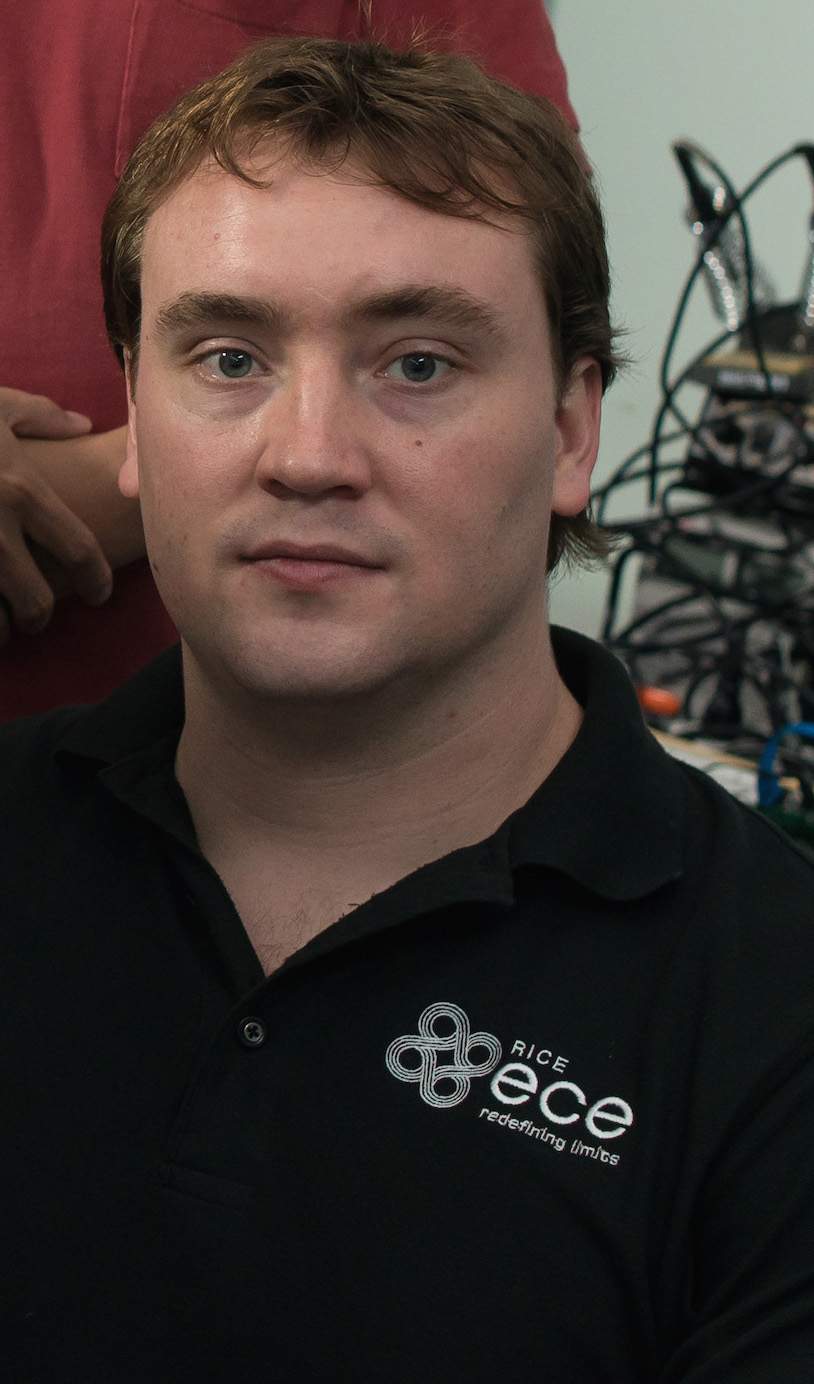}}]{Clayton Shepard} is a Ph.D. candidate at Rice University, where he also completed his B.S. in 2008, and M.S. in 2012.  He is a member of the Rice Efficient Computing Group, lead by Dr. Lin Zhong.  In 2008 he was a visiting researcher with Motorola's Advanced Research and Technology lab, where he also interned in 2007.  His research interests include mobile systems and ubiquitous low-power computing, his current research focus is many-antenna base stations.  In 2011 and 2012 he interned with Bell Labs, Alcatel-Lucent.  He received the NDSEG fellowship award in 2011.
\end{IEEEbiography}

\begin{IEEEbiography}[{\includegraphics[width=1in,height=1.25in,clip,keepaspectratio]{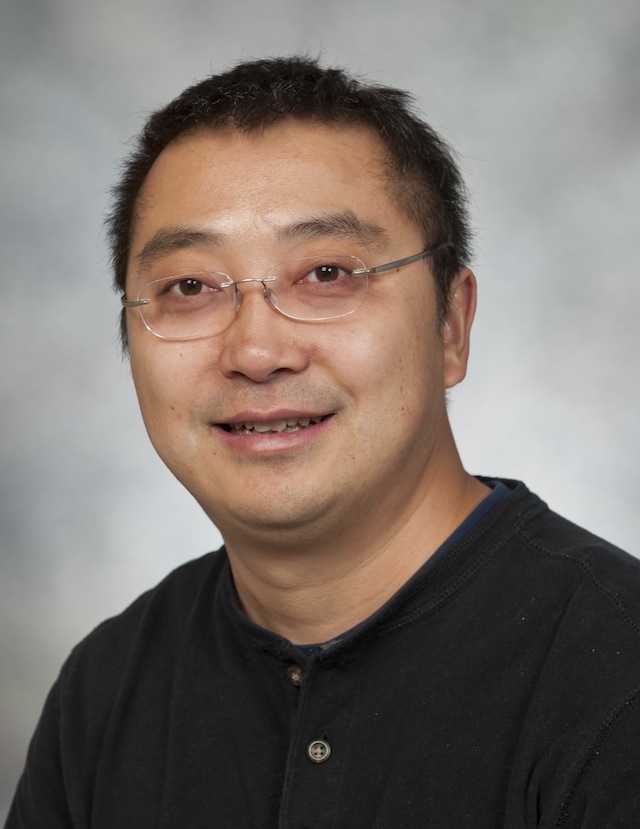}}]{Lin Zhong} received the BS and MS degrees from Tsinghua University in 1998 and 2000, respectively, and the PhD degree from Princeton University in September 2005. He is a Professor in the Department of Electrical and Computer Engineering, Rice University. His work received the Best Paper Awards from ACM MobileHCI 2007, IEEE PerCom 2009, ACM MobiSys 2011, 2013 and 2014, and ACM ASPLOS 2014. A paper he coauthored was recognized as one of the 30 most influential papers in the first 10 years of the Design, Automation and Test in Europe conference. He is a recipient of of the US National Science Foundation CAREER award, the Duncan Award from Rice University and the Rockstar Award from ACM SIGMOBILE. His research interests include mobile  and wireless systems. He is a senior member of the IEEE.\end{IEEEbiography}

\begin{IEEEbiography}[{\includegraphics[width=1in,height=1.25in,clip,keepaspectratio]{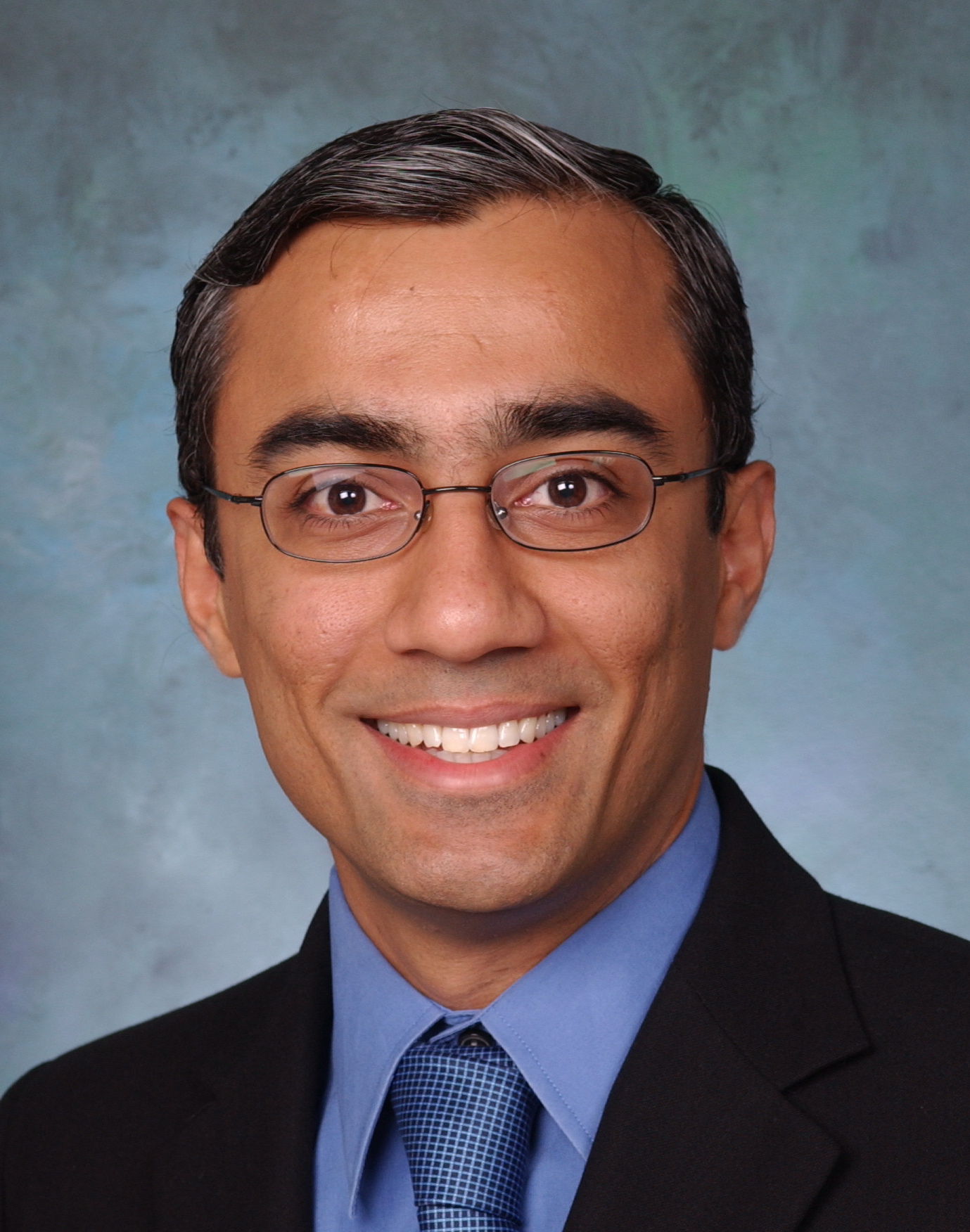}}]{Ashutosh Sabharwal} received the B. Tech. degree from the Indian Institute of Technology, New Delhi, India, in 1993 and the M.S. and Ph.D. degrees from The Ohio State University, Columbus, in 1995 and 1999, respectively. He is currently a Professor with the Department of Electrical and Computer Engineering, Rice University, Houston, TX. His research interests include information theory, communication algorithms and experiment-driven design of wireless networks. He received the 1998 Presidential Dissertation Fellowship Award. \end{IEEEbiography}

\end{document}